\documentclass[my]{myour}
\usepackage{latexsym}
\usepackage{amssymb}
\usepackage{amsmath}
\usepackage{graphics}
\usepackage{epsf}
\usepackage[dvips]{graphicx}
\begin{document}
\title{Energy dependence of slope parameter in elastic nucleon-nucleon scattering}

\author{V.A. Okorokov\inst{}
\thanks{e-mail: VAOkorokov@mephi.ru;~Okorokov@bnl.gov}%
} 
\institute{National Research Nuclear University ``MEPhI" (Moscow
Engineering Physics Institute), \\ Kashirskoe Shosse 31, 115409
Moscow, Russia}
\date{\today}
\abstract{The study of slope parameter is presented for elastic
proton-proton and antiproton-proton scattering with taking into
account the resent experimental data at high energies. The
expanded logarithmic approximations allow the description of the
experimental slopes in all available energy range reasonably.
Accounting for the LHC results leads to the dramatic change of
behavior of the quadratic in logarithm approximation at high
energies and to the closer trends for all fitting functions under
study in comparison with the analysis at collision energies up to
the 200 GeV. The estimations of the asymptotic shrinkage parameter
$\alpha'_{\cal{P}}$ are discussed. Predictions for diffraction
slope parameter are obtained for some proton-proton and
antiproton-proton facilities.  
} 
\maketitle
\section{Introduction}\label{intro}
The one of the important quantity for nucleon elastic scattering
is the slope parameter $B$ which is defined in accordance with the
following equation:
\begin{equation}
B(s,t)=\partial_{t}\ln \partial_{t} \sigma(s,t),
\label{eq:Slope-def}
\end{equation}
where $\partial_{t} \equiv \partial / \partial t$. The $B$ is
determined experimentally by fitting the differential cross
section $d\sigma / dt$ at some collision energy $\sqrt{s}$. The
study of $B$ parameter is important, in particular, for
reconstruction procedure of full set of helicity amplitudes for
elastic nucleon-nucleon scattering \cite{Okorokov-SPIN-2007}. In
the last 20--30 years, high-energy $\bar{p}p$ colliders have
extended the maximum $\bar{p}p$ collision energy from $\sqrt{s}
\sim 20$ GeV to $\sqrt{s} \sim 2$ TeV, the LHC facility allows one
to obtain $pp$ data up to $\sqrt{s}=8$ TeV so far. As consequence,
the available collection of $pp$ and $\bar{p}p$ slope data from
literature has extended. The $B(s)$ for elastic $pp$ and
$\bar{p}p$ reactions is under consideration below. Within the
classical Regge model the Pomeron trajectory,
$\alpha_{\cal{P}}(t)$, is linear function of momentum transfer,
i.e. $\alpha_{\cal{P}}(t)=\alpha_{\cal{P}}(0)+\alpha'_{\cal{P}}t$.
Therefore using the definition (\ref{eq:Slope-def}) the following
relation can be obtained for $B(s)$ at some fixed $t$:
\begin{equation*}
B(s) \propto 2\alpha'_{\cal{P}}\ln \varepsilon,
\end{equation*}
where $\varepsilon \equiv s/s_{0}$, $s_{0}=1$ GeV$^{2}$. In
general case for Pomeron-inspired models the asymptotic shrinkage
parameter $\alpha'_{\cal{P}}$ can be written as follows:
$2\alpha'_{\cal{P}}(s)=\partial_{\ln \varepsilon} B(s,t)$.
\section{Experimental slope energy dependence}\label{sec:1}
As in the previous analyses
\cite{Okorokov-arXiv-0811.0895,Okorokov-arXiv-0811.3849,Okorokov-arXiv-0907.0951}
dependence $B(s)$ is parameterized by the following analytic
functions:
\begin{subequations}
\begin{eqnarray}
B(s,t)&=& B_{0}+2a_{1}\ln \varepsilon,
\label{eq:Fit-1} \\
B(s,t)&=& B_{0}+2a_{1}\ln \varepsilon +a_{2}\ln^{a_{3}}
\varepsilon,
\label{eq:Fit-2}\\
B(s,t)&=& B_{0}+2a_{1}\ln \varepsilon +a_{2}\varepsilon^{a_{3}},
\label{eq:Fit-3}\\
B(s,t)&=& B_{0}+2a_{1}\ln \varepsilon +a_{2} \ln^{2} \varepsilon,
\label{eq:Fit-4}
\end{eqnarray}
\end{subequations}
where free parameters $B_{0},~ a_{i}, i=1-3$ depend on range of
$|t|$ which is used for approximation. There are the relations
$\alpha'_{\cal{P}}=a_{1}$ and
$\alpha'_{\cal{P}}=a_{1}+a_{2}\ln\varepsilon$ for
parameterizations (\ref{eq:Fit-1}) and (\ref{eq:Fit-4}) inspired
by the Pomeron exchange models. Appendix shows database of
available experimental results for slope parameter in elastic $pp
/ \bar{p}p$ scattering. The some numerical values of the $B(s,t)$
from this database are used in the present study. Experimental
values of slope parameter collected at initial energies $\sqrt{s}
\leq 1.8$ TeV were used in \cite{Okorokov-arXiv-0907.0951}.
Additional experimental results from Tevatron and the LHC are from
\cite{Abazov-PRD-86-012009-2012} and
\cite{Antchev-LJEFP-101-21002-2013,Aad-NPB-889-486-2014,Antchev-PRL-111-021001-2013,Antchev-LJEFP-95-41001-2011},
respectively. The full data sample consists of 490 experimental
points. The number of experimental points is equal $145 / 138$
($137 / 70$) for $pp / \bar{p}p$ scattering at low (intermediate)
$|t|$ respectively. It should be noted that for intermediate $|t|$
range the experimental results for $B(s)$ obtained with help of
linear parametrization for logarithm of differential
cross-section, $\ln\left(d\sigma/dt\right) \propto
\left(-B|t|\right)$, are discussed below because the new
experimental data at intermediate $|t|$
\cite{Antchev-LJEFP-95-41001-2011,Abazov-PRD-86-012009-2012} with
respect to the \cite{Okorokov-arXiv-0907.0951} are obtained for
such parametrization of $\ln\left(d\sigma/dt\right)$ namely. From
the exponential parametrization with index quadratic in $t$ for
differential cross-section, $\ln\left(d\sigma/dt\right) \propto
\left(-B|t| \pm Ct^{2}\right)$, one may calculate the local slope
as following
\begin{equation}
\left.b\left(s,t\right)\right|_{|t|=0.2
\scriptsize{\mbox{GeV}}^{2}}=B \pm C\ln|t|,~~~ B,C > 0.
\label{eq:b-Def}
\end{equation}
In accordance with \cite{Okorokov-arXiv-0907.0951} the mean value
of $|t|~(|\bar{t}|)$ is calculated with taking into account the
approximation of experimental $d\sigma/dt$ distribution; errors of
experimental points include available clear indicated systematic
errors added in quadrature to statistical ones. The fitting
algorithm is described in detail in
\cite{Okorokov-arXiv-0907.0951}. As previously the points with
$n_{\chi} \geq 2$ are excluded from fit in our algorithm, where
\cite{Okorokov-arXiv-0907.0951}
\begin{equation}
n_{\chi}=\left(\frac{\textstyle
B_{m}^{i}-B\left(s_{i};\vec{\alpha}\right)}{\textstyle
\sigma_{i}\sqrt{\chi^{2}/\mbox{n.d.f.}}}\right)^{2}.
\label{eq:Chi-rel}
\end{equation}
Here $B_{m}^{i}$ is the measured value of nuclear slope at $s_{i}$
with experimental error $\sigma_{i}$,
$B\left(s_{i};\vec{a}\right)$ is the expected value from the
fitting function with best $\chi^{2}/\mbox{n.d.f.}$ among
(\ref{eq:Fit-1}) -- (\ref{eq:Fit-4}) for approximation of all
range of available energies, the parameters $\alpha_{j}$ are from
$N$-dimensional vector
$\vec{\alpha}=\left\{\alpha_{1},...,\alpha_{N}\right\}$. We
consider the estimates of fit parameters as the final results if
there are no excluded points for present data sample. The fraction
of excluded points is about $2\%$ for $pp$ as well as for
$\bar{p}p$ elastic scattering for low $|t|$ domain. The maximum
relative amount of rejected points is about $3\% / 12\%$ for
linear $\ln d\sigma / dt$ parametrization at intermediate $|t|$
values for $pp / \bar{p}p$ scattering respectively.
\subsection{Low $|t|$ domain}\label{sec:2.1}
\begin{figure}[h!]
\resizebox{0.5\textwidth}{!}{%
  \includegraphics[width=8.0cm,height=8.0cm]{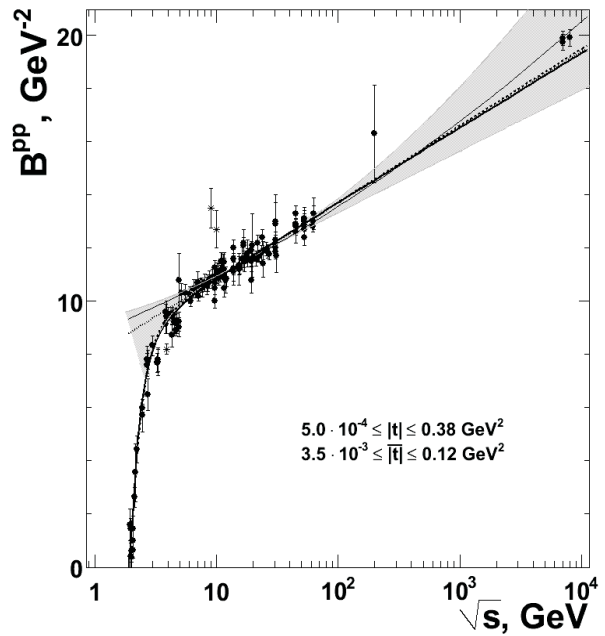}
} \caption{Energy dependence of the slope parameter for elastic
$pp$ scattering for low $|t|$ domain. Experimental points from
fitted samples are indicated as {\large$\bullet$}, unfitted points
are indicated as {\large$*$}. The dot curve is the fit of
experimental slope by the function (\ref{eq:Fit-1}), thick solid
-- by the (\ref{eq:Fit-2}), dashed -- by the (\ref{eq:Fit-3}),
thin solid -- by the (\ref{eq:Fit-4}). The shaded band corresponds
to the spread of fitting functions for previous analysis
\cite{Okorokov-arXiv-0907.0951}.} \label{fig:1}
\end{figure}

The experimental dependence $B(s)$ and corresponding fits by
(\ref{eq:Fit-1}) -- (\ref{eq:Fit-4}) are shown in Fig.\ref{fig:1}
for elastic $pp$ scattering, Table \ref{tab:1} contains values of
fit parameters. As seen the fitting functions (\ref{eq:Fit-1}),
(\ref{eq:Fit-4}) describe the $pp$ experimental data statistically
acceptable only for $\sqrt{s} \geq 5$ GeV. The RHIC point at
$\sqrt{s}=200$ GeV does not contradict the common trend within a
large error bars and can't discriminate the approximations under
study. In general the LHC results
\cite{Antchev-LJEFP-101-21002-2013,Antchev-PRL-111-021001-2013,Aad-NPB-889-486-2014}
added in fitting sample result in better agreement of fits
(\ref{eq:Fit-1}) -- (\ref{eq:Fit-4}) in comparison with previous
study \cite{Okorokov-arXiv-0907.0951}. All fits are very close to
each other in energy domain 10 GeV $\lesssim \sqrt{s} \lesssim$ 1
TeV, the quadratic in logarithm function (\ref{eq:Fit-4}) shows
some faster increasing of slope and noticeable difference from
another fit functions in multi-TeV region only. It seems the
ultra-high energy domain is suitable for separation of various
parameterizations. Fitting functions (\ref{eq:Fit-2}),
(\ref{eq:Fit-3}) allow us to describe experimental data at all
energies with reasonable fit quality for $pp$ (Table \ref{tab:1}).
The functions (\ref{eq:Fit-1}) -- (\ref{eq:Fit-3}) agrees very
well for $\sqrt{s} \geq 5$ GeV, furthermore there is no visible
difference between modifications (\ref{eq:Fit-2}) and
(\ref{eq:Fit-3}) in all experimentally available energy domain.
The function (\ref{eq:Fit-3}) demonstrate some better quality for
fit of full data sample (Table \ref{tab:1}) than function
(\ref{eq:Fit-2}) in contrast with previous analysis
\cite{Okorokov-arXiv-0907.0951}. The mean value of
(\ref{eq:Chi-rel}) for excluded points $(\overline{n_{\chi}}\,)$
is equal 4.7 for $pp$ data sample for parametrization (2c). The
accounting for LHC data leads to some decreasing of values of
$B_{0}$ and increasing of $a_{1}$ parameters for all fitting
functions (\ref{eq:Fit-1}) -- (\ref{eq:Fit-4}) under study in
comparison with values of corresponding parameters in previous
investigation \cite{Okorokov-arXiv-0907.0951}. This behavior of
$a_{1}$ with collision energy agrees well with predicted growth of
$\alpha'_{\cal{P}}$ with increasing of $\sqrt{s}$
\cite{Abe-PRD-50-5518-1994,Schegelsky-PRD-85-094024-2012}.
As seen from Table \ref{tab:1} the third term in both the
(\ref{eq:Fit-2}) and the (\ref{eq:Fit-3}) gives the main
contribution at $\sqrt{s} < 5$ GeV in the case of elastic
proton-proton scattering, i.e. describe the sharp changing of
slope in the low energy domain. Therefore $B^{pp} \propto \ln
\varepsilon$ at high $\sqrt{s}$ in accordance with
(\ref{eq:Fit-2}) and (\ref{eq:Fit-3}). Such asymptotic behavior is
in qualitative agreement with energy dependence of the slope
parameter for $\bar{p}p$ collisions
\cite{Okorokov-arXiv-0907.0951}.
The increasing of $a_{1}$ exhibits that $B^{pp}$ growth some
faster in multi-TeV region than one can expect from the trend
based on data sample at $\sqrt{s} \leq 200$ GeV. This suggestion
is confirmed by improvement of fit quality for one fitting
function (\ref{eq:Fit-4}) for present data sample in comparison
with fit qualities for experimental points at $\sqrt{s} \leq 200$
GeV \cite{Okorokov-arXiv-0907.0951}. Values of $a_{2}$ obtained in
present study and for fit at $\sqrt{s} \leq 200$ GeV are close
within errors for functions (\ref{eq:Fit-2}) and (\ref{eq:Fit-4})
but accounting for the LHC data results in some increasing of the
absolute value of $a_{3}$ in fit by (\ref{eq:Fit-2}). The absolute
values of $a_{2}$ and $a_{3}$ are the same within error bars for
function (\ref{eq:Fit-3}) for Table \ref{tab:1} and for fit in
energy domain $\sqrt{s} \leq 200$ GeV
\cite{Okorokov-arXiv-0907.0951}.

\begin{table*}
\caption{Values of parameters for fitting of slope energy
dependence in $pp$ elastic scattering for low $|t|$ domain}
\label{tab:1}
\begin{center}
\begin{tabular}{lccccc}
\hline \multicolumn{1}{l}{Function} &
\multicolumn{5}{c}{Parameter} \\
\cline{2-6} \rule{0pt}{10pt}
 & $B_{0}$, GeV$^{-2}$ & $a_{1}$, GeV$^{-2}$ & $a_{2}$, GeV$^{-2}$ & $a_{3}$ & $\chi^{2}/\mbox{n.d.f.}$ \\
\hline
(\ref{eq:Fit-1}) & $8.00 \pm 0.06$  & $0.309 \pm 0.010$ & --               & --               & $234/98$ \\
(\ref{eq:Fit-2}) & $8.09 \pm 0.06$  & $0.305 \pm 0.005$ & $-31.8 \pm 1.6$  & $-4.06 \pm 0.13$ & $420/138$ \\
(\ref{eq:Fit-3}) & $7.95 \pm 0.06$  & $0.313 \pm 0.005$ & $-240 \pm 32$    & $-2.23 \pm 0.09$ & $402/138$ \\
(\ref{eq:Fit-4}) & $8.81 \pm 0.12$  & $0.198 \pm 0.015$ & $0.013 \pm 0.002$& --               & $174/97$ \\
\hline
\end{tabular}
\end{center}
\end{table*}

\begin{table*}
\caption{Predictions for slope in nucleon-nucleon elastic
scattering at intermediate energies for low $|t|$ domain}
\label{tab:2a}
\begin{center}
\begin{tabular}{lcccccc}
\hline \multicolumn{1}{l}{Fitting} &
\multicolumn{6}{c}{Facility energies $\sqrt{s}$, GeV} \\
\cline{2-7} \rule{0pt}{10pt} function & \multicolumn{4}{c}{FAIR} &
\multicolumn{2}{c}{NICA}\\
\cline{2-7} \rule{0pt}{10pt}
 & 3 & 5 & 6.5 & 14.7 & 20 & 25\\
\hline
(\ref{eq:Fit-1}) & --               & $11.4 \pm 0.2$   & $11.6 \pm 0.2$   & $12.3 \pm 0.2$   & $11.70 \pm 0.13$ & $11.98 \pm 0.14$\\
(\ref{eq:Fit-2}) & $12.57 \pm 0.07$ & $12.80 \pm 0.07$ & $12.93 \pm 0.07$ & $13.32 \pm 0.09$ & $11.72 \pm 0.08$ & $12.00 \pm 0.09$\\
(\ref{eq:Fit-3}) & $12.59 \pm 0.05$ & $12.83 \pm 0.06$ & $12.95 \pm 0.06$ & $13.34 \pm 0.07$ & $11.70 \pm 0.08$ & $11.98 \pm 0.09$\\
(\ref{eq:Fit-4}) & --               & $11.9 \pm 1.2$   & $12.0 \pm 1.3$   & $12.4 \pm 1.6$   & $11.6 \pm 0.2$   & $11.9 \pm 0.2$\\
\hline
\end{tabular}
\end{center}
\end{table*}

\begin{table*}
\caption{Predictions for slope in $pp$ elastic scattering at high
energies for low $|t|$ domain} \label{tab:2b}
\begin{center}
\begin{tabular}{lccccccc}
\hline \multicolumn{1}{l}{Fitting} &
\multicolumn{7}{c}{Facility energies $\sqrt{s}$, TeV} \\
\cline{2-8} \rule{0pt}{10pt} function & \multicolumn{1}{c}{RHIC} &
\multicolumn{3}{c}{LHC} &
\multicolumn{2}{c}{FCC/VLHC} & \multicolumn{1}{c}{} \\
\cline{2-8} \rule{0pt}{10pt}
 & 0.5 & 14 & 28 & 42$^*$ & 100 & 200 & 500\\
\hline
(\ref{eq:Fit-1}) & $15.7 \pm 0.3$   & $19.8 \pm 0.4$ & $20.7 \pm 0.4$ & $21.2 \pm 0.4$ & $22.2 \pm 0.5$ & $23.1 \pm 0.5$ & $24.2 \pm 0.5$ \\
(\ref{eq:Fit-2}) & $15.67 \pm 0.14$ & $19.7 \pm 0.2$ & $20.6 \pm 0.2$ & $21.1 \pm 0.2$ & $22.1 \pm 0.2$ & $23.0 \pm 0.3$ & $24.1 \pm 0.3$ \\
(\ref{eq:Fit-3}) & $15.73 \pm 0.14$ & $19.9 \pm 0.2$ & $20.8 \pm 0.2$ & $21.3 \pm 0.2$ & $22.4 \pm 0.2$ & $23.2 \pm 0.3$ & $24.4 \pm 0.3$ \\
(\ref{eq:Fit-4}) & $15.7 \pm 0.5$   & $21.1 \pm 0.9$ & $22.4 \pm 1.0$ & $23.1 \pm 1.1$ & $24.8 \pm 1.3$ & $26.2 \pm 1.4$ & $28.2 \pm 1.6$ \\
\hline\multicolumn{8}{l}{$^*$\rule{0pt}{11pt}\footnotesize The
ultimate energy upgrade of LHC project \cite{Skrinsky-ICHEP2006}.}
\end{tabular}
\end{center}
\end{table*}
The comparison with earlier studies
\cite{Giacomelli-PR-23-123-1976,Burq-NPB-217-285-1983} confirms
the conclusion above that accounting for high energy data points
leads to increasing of value of $a_{1}^{pp}$ parameter and growth
of slope parameter seems to be faster in TeV-region than that at
lower energies. Accordingly the $a_{1}^{pp}$ value for function
(\ref{eq:Fit-4}) is closer significantly to the
$\alpha'_{\cal{P}}\approx 0.25$ GeV$^{-2}$ than that for previous
analysis in energy range $\sqrt{s} \leq 200$ GeV
\cite{Okorokov-arXiv-0907.0951}. On the other hand the values of
the $a_{1}^{pp}$ parameter obtained for fitting function
(\ref{eq:Fit-1}) is some larger than the prediction for
$\alpha'_{\cal{P}}$ from Pomeron inspired model for TeV-energy
domain \cite{Schegelsky-PRD-85-094024-2012}. Furthermore the
proton-proton results from fit by function (\ref{eq:Fit-4}) allow
the estimation $\left.2\alpha'_{\cal{P}}(s)\right|_{\sqrt{s}=8
\scriptsize{\mbox{TeV}}}=0.86 \pm 0.08$ which is almost twice
larger than the corresponding prediction from
\cite{Schegelsky-PRD-85-094024-2012}.

Predictions for $B$ are obtained for some facilities based on the
fit results shown above for $pp$ and on the results from
\cite{Okorokov-arXiv-0907.0951} for $\bar{p}p$. Estimations at low
$|t|$ for different intermediate energies of the projects FAIR and
NICA are shown in the Table \ref{tab:2a} and for high energy
domain are presented in the Table \ref{tab:2b}. As expected the
functions (\ref{eq:Fit-2}) and (\ref{eq:Fit-3}) predicts for FAIR
the $B$ values coincide with each other within errors. The
approximation functions (\ref{eq:Fit-1}) and (\ref{eq:Fit-4}) can
predict for $\sqrt{s} \geq 5$ GeV only. The Pomeron inspired
function (\ref{eq:Fit-1}) predicts slope parameter values smaller
significantly than that for modified fitting functions
(\ref{eq:Fit-2}) and (\ref{eq:Fit-3}) at FAIR energies. For
$\bar{p}p$ the predictions with help of quadratic in $\ln
\varepsilon$ function (\ref{eq:Fit-4}) are equal with estimations
based on any another fitting function under study within large
error bars at $\sqrt{s} \leq 14.7$ GeV. In the case of elastic
$pp$ scattering the functions (\ref{eq:Fit-1}) -- (\ref{eq:Fit-3})
predict just the equal values of $B$ within error bars at any
collision energy $\sqrt{s}$ discussed here. The difference between
estimations from (\ref{eq:Fit-1}) -- (\ref{eq:Fit-3}) and from
(\ref{eq:Fit-4}) onsets at the final energy of the LHC project
$\sqrt{s}=14$ TeV only. Our prediction with (\ref{eq:Fit-4})
function for RHIC energy is equal with early prediction for close
energy based only on slope data in the region $5 < \sqrt{s} < 62$
GeV \cite{Block-RevModPhys-57-563-1985} within errors. But
(\ref{eq:Fit-4}) underestimates the $B$ values in ultra-high
energy domain $\sqrt{s} > 40$ TeV in comparison with results based
on the approach without odderons
\cite{Block-RevModPhys-57-563-1985}. It should be emphasized that
in contrast with previous analysis \cite{Okorokov-arXiv-0907.0951}
the present fits by functions (\ref{eq:Fit-1}) -- (\ref{eq:Fit-4})
of data sample included of the LHC results predict the similar
increasing of $B$ with energy as most of phenomenological models
\cite{Kundrat-EDS-273-2007}. The $B$ value predicted for the LHC
at $\sqrt{s}=14$ TeV by (\ref{eq:Fit-1}) only is close with errors
to the predictions from
\cite{Block-PRD-60-054024-1999,Petrov-EPJ-C28-525-2003}, in
particular, with estimation from model with hadronic amplitude
corresponding to the exchange of two pomerons. Prediction of
phenomenological model with hadronic amplitude corresponding to
the exchange of three pomerons \cite{Petrov-EPJ-C28-525-2003} at
$\sqrt{s}=14$ TeV coincides with estimation of $B$ within error
bars from fit function (\ref{eq:Fit-4}) with fastest growth of $B$
with $\sqrt{s}$ in multi-TeV region. But most of estimations of
$B$ at $\sqrt{s}=14$ TeV from Table \ref{tab:2b} agree well within
errors with model prediction from \cite{Bourrely-EPJ-C28-97-2003}.
However the model estimates at $\sqrt{s}=14$ TeV described above
were obtained for $B\left(t=0\right)$ and the $t$-dependence of
slope shows the slight decreasing of $B$ for the model with
three-pomeron exchange \cite{Petrov-EPJ-C28-525-2003} and faster
decreasing of $B$ for the model from
\cite{Bourrely-EPJ-C28-97-2003} at growth of momentum transfer up
to $|t| \approx 0.1$ GeV$^{2}$. Therefore one can expect that the
model with hadronic amplitude corresponding to the exchange of
three pomerons \cite{Petrov-EPJ-C28-525-2003} will be in the
better agreement with values of $B$ from Table \ref{tab:2b}
predicted for finite (non-zero) low $|t|$ values. But the model
from \cite{Islam-PLB-605-115-2005} overestimates the $B$ value at
$\sqrt{s}=14$ TeV in comparison with corresponding predictions
from fitting functions (\ref{eq:Fit-1}) -- (\ref{eq:Fit-4})
despite of the sharp decreasing of $B$ at growth of momentum
transfer up to $|t| \approx 0.1$ GeV$^{2}$ in this model. As
suggested sometimes the saturation regime, Black Disk Limit (BDL),
maybe reach at the LHC. The one of the models in which such
effects appear, namely, Dubna Dynamical Model (DDM) predicts the
slope $B\left(t=0\right) \approx 23.5$ GeV$^{-2}$ at $\sqrt{s}=14$
TeV \cite{Selyugin-EDS2007-279} which is larger noticeably than
the predictions from Table \ref{tab:2b} at the same $\sqrt{s}$.
Therefore the saturation regime will not be reached, at least, at
the LHC energy $\sqrt{s}=14$ TeV as suggested, for example, in the
model from \cite{Okorokov-IJMPA-25-5333-2010} or simple saturation
can not be enough in order to describe the LHC data at
quantitative level.
\subsection{Intermediate $|t|$ domain}
\label{sec:2.1} 
\begin{figure*}
\begin{center}
\begin{tabular}{cc}
\mbox{
\resizebox{0.5\textwidth}{!}{%
  \includegraphics[width=8.0cm,height=8.0cm]{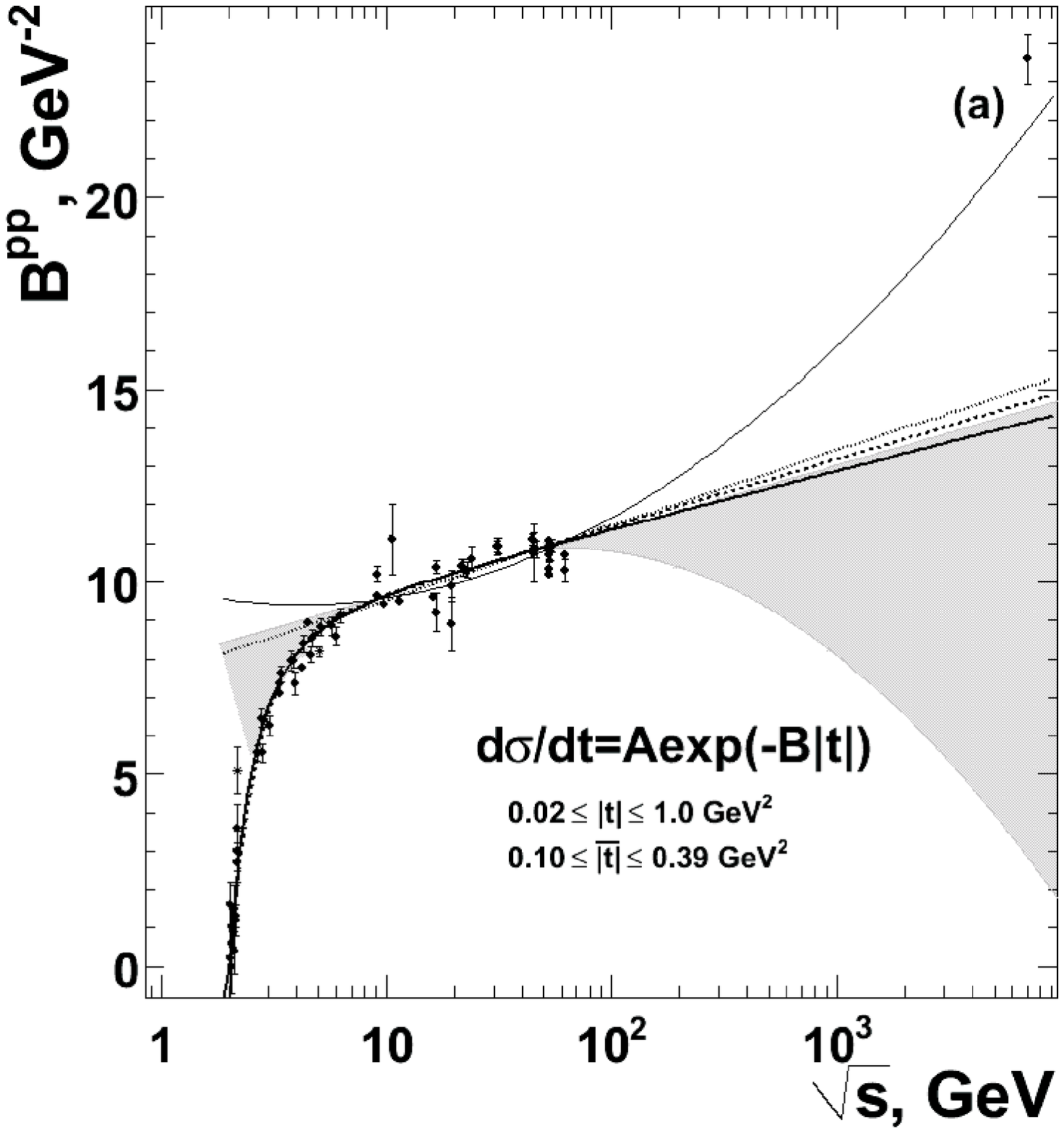}
}} & \mbox{
\resizebox{0.5\textwidth}{!}{%
  \includegraphics[width=8.0cm,height=8.0cm]{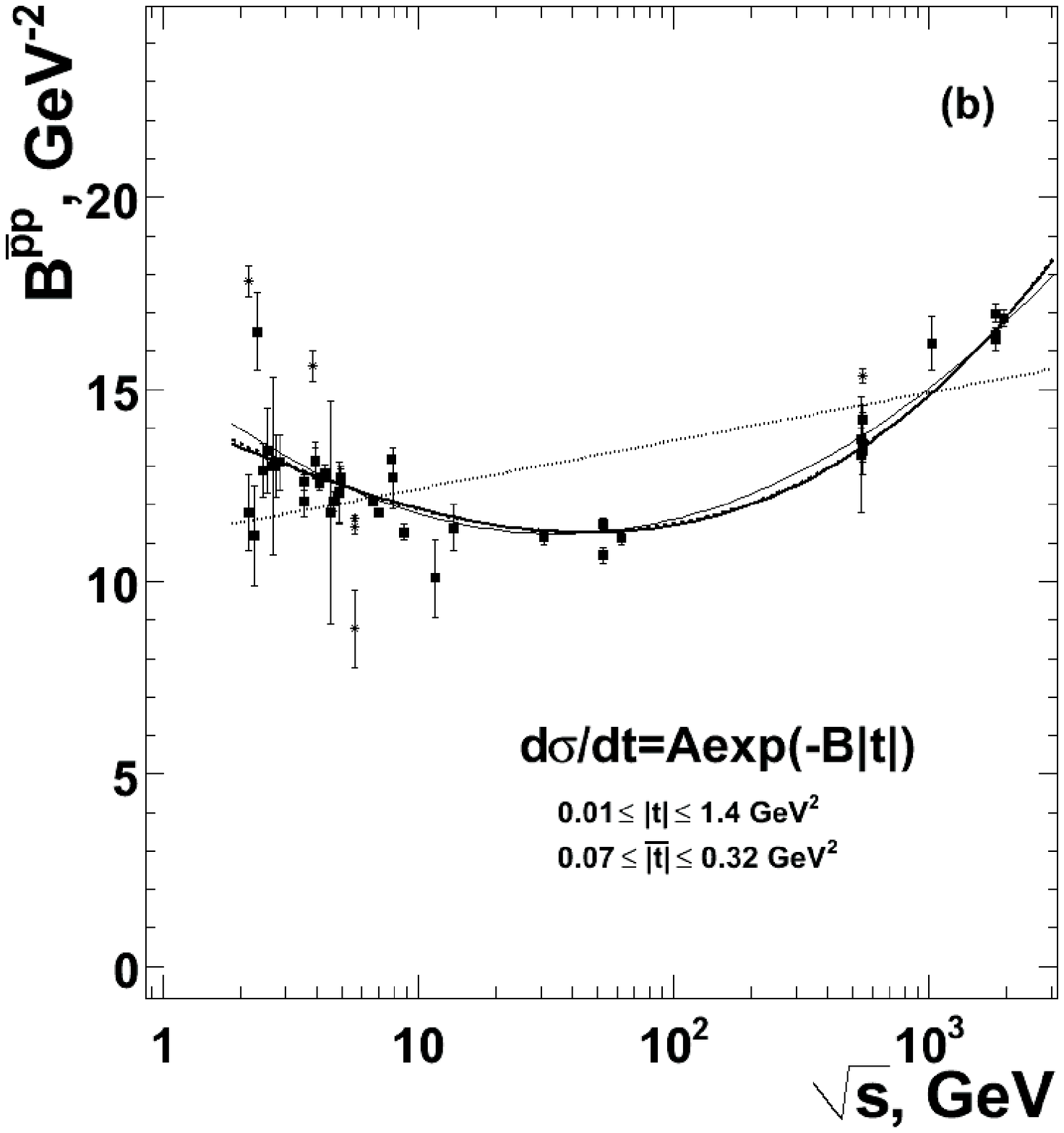}
}}
\end{tabular}
\end{center}
\vspace*{-0.4cm} \caption{Energy dependence of $B$ in
proton-proton (a) and antiproton-proton (b) elastic scattering for
exponential parametrization with index linear in $|t|$ of
differential cross section. Experimental points from fitted
samples are indicated as close circles / squares for $pp /
\bar{p}p$, unfitted points are indicated as {\large$*$}. The dot
curve is the fit of experimental slope by the function
(\ref{eq:Fit-1}), thick solid -- by the (\ref{eq:Fit-2}), dashed
-- by the (\ref{eq:Fit-3}), thin solid -- by the (\ref{eq:Fit-4}).
The shaded band (a) corresponds to the spread of fitting functions
for previous analysis
\cite{Okorokov-arXiv-0907.0951}.}\label{fig:3}
\end{figure*}

As indicated in \cite{Okorokov-arXiv-0907.0951}
the situation is more complicated for intermediate $|t|$ domain.
Figure \ref{fig:3} shows the experimental data and corresponding
fits for energy dependence of slope parameter at intermediate
$|t|$ for exponential approximation with index linear in $|t|$ of
differential cross-sections in elastic $pp$ and $\bar{p}p$
scattering. The parameter values for fitting are indicated in
Table \ref{tab:3} for various interaction types.

Usually the fit qualities are poorer for intermediate $|t|$ values
than that for low $|t|$ range in $pp$ elastic collisions for
linear parametrization of $\ln (d\sigma / dt)$. The fitting
functions (\ref{eq:Fit-1}) and (\ref{eq:Fit-4}) agree with
experimental points qualitatively for $\sqrt{s} \geq 5$ GeV only.
Furthermore in the first case there is significant discrepancy
between experimental point and fit curve at the LHC energy also
(Fig.\ref{fig:3}a). The ``expanded" functions (\ref{eq:Fit-2}),
(\ref{eq:Fit-3}) approximate experimental data at all energies
reasonably with close fit qualities (Table \ref{tab:3}), but these
functions show a slow growth of slope parameter with energy
increasing at $\sqrt{s} \geq 100$ GeV (Fig.\ref{fig:3}a). It
should be stressed that the experimental point at the LHC energy
leads to the dramatic change of behavior of the fitting function
(\ref{eq:Fit-4}) in comparison with previous analysis
\cite{Okorokov-arXiv-0907.0951}. At present the fitting function
(\ref{eq:Fit-4}) predicts increasing of the nuclear slope in high
energy domain as well as all other fitting functions under study.
Such behavior is opposite to the result of fit by function
(\ref{eq:Fit-4}) of experimental data sample at $\sqrt{s} \leq
200$ GeV \cite{Okorokov-arXiv-0907.0951}. As shown in previous
analysis \cite{Okorokov-arXiv-0907.0951} the the $\bar{p}p$
experimental data admit the approximation by (\ref{eq:Fit-1}),
(\ref{eq:Fit-4}) for all energy range but not only for $\sqrt{s}
\geq 5$ GeV (Fig.\ref{fig:3}b). Thus the parameter values are
shown in Table \ref{tab:3} for approximation by (\ref{eq:Fit-1}),
(\ref{eq:Fit-4}) of all available experimental data. The fit
curves show (very) close behaviors for both the present and
previous analyses in the case of the Fig.\ref{fig:3}b. The
$\bar{p}p$ data disagreement with Pomeron inspired fitting
function (\ref{eq:Fit-1}) very significantly (Fig.\ref{fig:3}b).
Functions (\ref{eq:Fit-2}) and (\ref{eq:Fit-3}) show a close
behavior at all energies for $\bar{p}p$ data from linear
parametrization of $\ln d\sigma/dt$. These fitting functions have
a better fit quality than (\ref{eq:Fit-4}) but fits by functions
(\ref{eq:Fit-2}), (\ref{eq:Fit-3}) are still statistically
unacceptable. The $\overline{n_{\chi}}=2.9$ for excluded $pp$ data
with (\ref{eq:Fit-3}) function and $\overline{n_{\chi}} = 18.3$
for points excluded from $\bar{p}p$ fitted data sample for
(\ref{eq:Fit-2}) fitting function. As seen from Fig.\ref{fig:3}b
behavior of the energy dependence of slope parameter for elastic
$\bar{p}p$ scattering is close to the quadratic in logarithm
function $B^{\bar{p}p} \propto \ln^{2} \varepsilon$ at high
$\sqrt{s}$ as well as for elastic $pp$ collisions. But in the last
case it is difficultly to make the unambiguous conclusion because
there is only the LHC point in the asymptotic region. The
estimations of asymptotic shrinkage parameter are following for
fitting functions (\ref{eq:Fit-1}) and (\ref{eq:Fit-4})
respectively: $2\alpha'_{\cal{P}}=0.422 \pm 0.016$ and
$\left.2\alpha'_{\cal{P}}(s)\right|_{\sqrt{s}=7
\scriptsize{\mbox{TeV}}}=1.66 \pm 0.19$ for $pp$;
$2\alpha'_{\cal{P}}=0.272 \pm 0.008$ and
$\left.2\alpha'_{\cal{P}}(s)\right|_{\sqrt{s}=1.96
\scriptsize{\mbox{TeV}}}=1.36 \pm 0.10$ for $\bar{p}p$. The values
of $\alpha'_{\cal{P}}$ are equal within errors at $\sqrt{s}=7$ TeV
and $\sqrt{s}=8$ TeV for domain of $|t|$ under consideration in
the case of the function (\ref{eq:Fit-4}) for $pp$ interaction. As
seen from comparison with results from Table \ref{tab:1} for
elastic $pp$ collisions and from \cite{Okorokov-arXiv-0907.0951}
for $\bar{p}p$ scattering the fitting function (\ref{eq:Fit-1}) is
characterized by smaller asymptotic shrinkage parameter for
intermediate $|t|$ domain than that at small $|t|$ values while
the situation for quadratic in logarithm function (\ref{eq:Fit-4})
is opposite: there is significant growth of $\alpha'_{\cal{P}}$
for transition from small $|t|$ to intermediate $|t|$ values.

\begin{table*}
\caption{Values of parameters for fitting of energy dependence of
slope at intermediate $|t|$} \label{tab:3}
\begin{center}
\begin{tabular}{lccccc}
\hline \multicolumn{1}{l}{Function} &
\multicolumn{5}{c}{Parameter} \\
\cline{2-6} \rule{0pt}{10pt}
 & $B_{0}$, GeV$^{-2}$ & $a_{1}$, GeV$^{-2}$ & $a_{2}$, GeV$^{-2}$ & $a_{3}$ & $\chi^{2}/\mbox{n.d.f.}$ \\
\hline
\multicolumn{6}{c}{proton-proton scattering, experimental data for $d\sigma/dt=A\exp\left(-B|t|\right)$} \\
\hline
(\ref{eq:Fit-1}) & $7.59 \pm 0.11$ & $0.211 \pm 0.008$ & --               & --               & $322/35$ \\
(\ref{eq:Fit-2}) & $8.39 \pm 0.17$ & $0.163 \pm 0.011$ & $-25.2 \pm 1.4$  & $-3.01 \pm 0.13$ & $493/61$ \\
(\ref{eq:Fit-3}) & $7.94 \pm 0.11$ & $0.19 \pm 0.09$   & $-90 \pm 8$      & $-1.69 \pm 0.06$ & $458/61$ \\
(\ref{eq:Fit-4}) & $9.9 \pm 0.2  $ & $-0.16 \pm 0.03$  & $0.056 \pm 0.005$& --               & $187/34$ \\
\hline
\multicolumn{6}{c}{antiproton-proton scattering, experimental data for $d\sigma/dt=A\exp\left(-B|t|\right)$} \\
\hline
(\ref{eq:Fit-1}) & $11.16 \pm 0.06$ & $0.136 \pm 0.004$  & --                  & --                  & $1168/42$ \\
(\ref{eq:Fit-2}) & $14.34 \pm 0.11$ & $-0.304 \pm 0.014$ & $0.0042 \pm 0.0004$ & $2.92 \pm 0.04$     & $186/40$ \\
(\ref{eq:Fit-3}) & $7.7 \pm 0.2$    & $-0.735 \pm 0.019$ & $6.9 \pm 0.2$       & $0.1000 \pm 0.0013$ & $188/40$ \\
(\ref{eq:Fit-4}) & $15.48 \pm 0.15$ & $-0.60 \pm 0.02$   & $0.084 \pm 0.003$   & --                  & $197/41$ \\
\hline
\end{tabular}
\end{center}
\vspace*{-0.4cm}
\end{table*}
\begin{table*}
\caption{Predictions for slopes in nucleon-nucleon elastic
scattering at intermediate energies for intermediate $|t|$ domain}
\label{tab:5}
\begin{center}
\begin{tabular}{lcccccc}
\hline \multicolumn{1}{l}{Fitting} &
\multicolumn{6}{c}{Facility energies $\sqrt{s}$, GeV} \\
\cline{2-7} \rule{0pt}{11pt} function & \multicolumn{4}{c}{FAIR} &
\multicolumn{2}{c}{NICA}\\
\cline{2-7} \rule{0pt}{10pt}
 & 3& 5 & 6.5 & 14.7 & 20 & 25 \\
\hline & \multicolumn{6}{c} {$B$-parameter} \\ \hline
(2a) & $11.76 \pm 0.06$ & $12.04 \pm 0.07$ & $12.18 \pm 0.07$ & $12.62 \pm 0.07$ & $10.12 \pm 0.15$ & $10.31 \pm 0.15$ \\
(2b) & $12.88 \pm 0.13$ & $12.51 \pm 0.14$ & $12.26 \pm 0.15$ & $11.6 \pm 0.2$   & $10.2 \pm 0.2$   & $10.4 \pm 0.2$\\
(2c) & $13.1 \pm 0.3$   & $12.5 \pm 0.4$   & $12.2 \pm 0.4$   & $11.6 \pm 0.5$   & $10.2 \pm 1.1$   & $10.4 \pm 1.2$ \\
(2d) & $13.27 \pm 0.18$ & $12.5 \pm 0.2$   & $12.2 \pm 0.2$   & $11.5 \pm 0.3$   & $10.0 \pm 0.5$   & $10.1 \pm 0.5$ \\
\hline & \multicolumn{6}{c} {$b$-parameter} \\ \hline
(2a) & $10.0 \pm 0.2$ & $10.5 \pm 0.2$ & $10.7 \pm 0.3$ & $11.4 \pm 0.3$ & $10.5 \pm 0.3$ & $10.8 \pm 0.4$ \\
(2b) & $14.3 \pm 1.4$ & $12.6 \pm 1.7$ & $12.0 \pm 1.8$ & $11 \pm 2$     & $10 \pm 2$     & $11 \pm 3$ \\
(2c) & $14 \pm 3$     & $13 \pm 3$     & $12 \pm 3$     & $11 \pm 3$     & $10.5 \pm 1.1$ & $10.7 \pm 1.2$ \\
(2d) & $12.6 \pm 1.1$ & $12.0 \pm 1.2$ & $11.8 \pm 1.3$ & $11.3 \pm 1.7$ & $10 \pm 2$     & $11 \pm 3$ \\
\hline
\end{tabular}
\end{center}
\vspace*{-0.4cm}
\end{table*}
\begin{table*}
\caption{Predictions for slopes in $pp$ elastic scattering at high
energies for intermediate $|t|$ domain} \label{tab:5dop}
\begin{center}
\begin{tabular}{lccccccc}
\hline \multicolumn{1}{l}{Fitting} &
\multicolumn{7}{c}{Facility energies $\sqrt{s}$, TeV} \\
\cline{2-8} \rule{0pt}{11pt} function & \multicolumn{1}{c}{RHIC} &
\multicolumn{3}{c}{LHC} &
\multicolumn{2}{c}{FCC/VLHC} & \\
\cline{2-8} \rule{0pt}{10pt}
 & 0.5& 14 & 28 & 42$^*$ & 100 & 200 & 500 \\
\hline & \multicolumn{7}{c} {$B$-parameter} \\ \hline
(2a) & $12.8 \pm 0.2$ & $15.6 \pm 0.3$ & $16.2 \pm 0.3$ & $16.6 \pm 0.4$ & $17.3 \pm 0.4$ & $17.9 \pm 0.4$ & $18.7 \pm 0.4$ \\
(2b) & $12.4 \pm 0.3$ & $14.6 \pm 0.4$ & $15.1 \pm 0.5$ & $15.3 \pm 0.5$ & $15.9 \pm 0.5$ & $16.3 \pm 0.6$ & $16.9 \pm 0.6$ \\
(2c) & $13 \pm 2$     & $15 \pm 3$     & $16 \pm 4$     & $16 \pm 4$     & $17 \pm 4$     & $17 \pm 4$     & $18 \pm 5$ \\
(2d) & $14.5 \pm 1.1$ & $24 \pm 2$     & $27 \pm 2$     & $28 \pm 3$     & $32 \pm 3$     & $35 \pm 3$     & $40 \pm 4$ \\
\hline & \multicolumn{7}{c} {$b$-parameter} \\ \hline
(2a) & $14.6 \pm 0.6$ & $18.8 \pm 0.9$ & $19.7 \pm 1.0$ & $20.2 \pm 1.0$ & $21.3 \pm 1.1$ & $22.2 \pm 1.1$ & $23.3 \pm 1.2$ \\
(2b) & $13 \pm 4$     & $16 \pm 5$     & $16 \pm 5$     & $17 \pm 5$     & $17 \pm 6$     & $18 \pm 6$     & $18 \pm 6$ \\
(2c) & $13.7 \pm 1.9$ & $17 \pm 3$     & $18 \pm 3$     & $18 \pm 3$     & $19 \pm 3$     & $20 \pm 3$     & $21 \pm 4$ \\
(2d) & $12 \pm 5$     & $9 \pm 9$      & $8 \pm 10$     & $7 \pm 11$     & $6 \pm 12$     & $4 \pm 13$     & $2 \pm 14$ \\
\hline \multicolumn{6}{l}{$^*$\rule{0pt}{11pt}\footnotesize The
ultimate energy upgrade of LHC project \cite{Skrinsky-ICHEP2006}.}
\end{tabular}
\end{center}
\vspace*{-0.4cm}
\end{table*}

We have obtained predictions for nuclear slope parameters $B$ and
$b$ for some facilities and intermediate $|t|$. The $B$ values are
calculated on the base of fit results shown above. Estimations for
$b$ are obtained on the base of results from
\cite{Okorokov-arXiv-0907.0951} with all functions under
consideration with exception of the (\ref{eq:Fit-2}). In the last
case the following results are obtained within framework of the
present analysis for $b(s)$: $B_{0}=9 \pm 2\,(50.8 \pm 0.9)$
GeV$^{-2}$, $a_{1}=0.19 \pm 0.12\,(0.92 \pm 0.06)$ GeV$^{-2}$,
$a_{2}=-10 \pm 2\,(-34.1 \pm 0.8)$ GeV$^{-2}$, $a_{3}=-1.9 \pm
0.5\,(0.222 \pm 0.016)$,
$\chi^{2}/\mbox{n.d.f.}=41.4/33\,(8.7/15)$ for $pp\,(\bar{p}p)$
data samples, respectively. Slope values are shown in the Table
\ref{tab:5} for different energies of FAIR, NICA, and in the Table
\ref{tab:5dop} for RHIC, LHC and FCC/VLHC. According to the fit
range function (\ref{eq:Fit-1}) can predicts the $B$ value for
$\bar{p}p$ scattering at all energies under study not in $\sqrt{s}
\geq 5$ GeV domain only. As expected the functions
(\ref{eq:Fit-2}) -- (\ref{eq:Fit-4}) predicted very close values
for slope parameter $B$ for FAIR. These fitting functions,
especially (\ref{eq:Fit-2}) and (\ref{eq:Fit-3}), predict the
close values for nuclear slope $B$ in NICA energy domain too.
Functions (\ref{eq:Fit-1}) -- (\ref{eq:Fit-3}) predict smaller
values for $B$ in high-energy $pp$ collisions than
(\ref{eq:Fit-4}) approximation especially for FCC/VLHC energy
domain. Perhaps, the future more precise RHIC results will be
useful for discrimination of fitting functions under study for
intermediate $|t|$ values. In the contrast with previous analysis
\cite{Okorokov-arXiv-0907.0951}, here the function
(\ref{eq:Fit-4}) with obtained parameters predicts fast growth of
$B$ values at energies of future experiments. This behavior of
estimations calculated for functions (\ref{eq:Fit-1}) --
(\ref{eq:Fit-4}) contradicts with earlier predictions from some
phenomenological models. It should be emphasized that various
phenomenological models predict a very sharp decreasing of nuclear
slope in the range $|t| \sim 0.3 - 0.5$ GeV$^{2}$ at LHC energy
$\sqrt{s}=14$ TeV \cite{Kundrat-EDS-273-2007}. Just the positive
$B$ value predicted for LHC at $\sqrt{s}=14$ TeV by
(\ref{eq:Fit-4}) is most close to the some model expectations
\cite{Petrov-EPJ-C28-525-2003,Bourrely-EPJ-C28-97-2003}. Taking
into account predictions in Table \ref{tab:2a} based on the
fitting functions (\ref{eq:Fit-1}) -- (\ref{eq:Fit-4}) for low
$|t|$ one can suggest that the model with hadronic amplitude
corresponding to the exchange of three pomerons
\cite{Kundrat-EDS-273-2007,Petrov-EPJ-C28-525-2003} describes the
nuclear slope some closer to the experimentally inspired values at
LHC energy both at low and intermediate $|t|$ than other models.
The situation with predictions for $b$ at intermediate energies is
similar to that for $B$: functions (\ref{eq:Fit-2}) --
(\ref{eq:Fit-4}) predict close values for $b$ within large errors
for both the FAIR and the NICA. Furthermore all fitting functions
predict the close values of $b$ at $\sqrt{s} > 5$ GeV. The value
of $b$ parameter obtained from function (\ref{eq:Fit-1}) differs
significantly from estimations with other fitting functions for
$\bar{p}p$ elastic scattering at low energies. Therefore the $b$
parameter seems more perspective for distinguishing of Pomeron
inspired function (\ref{eq:Fit-1}) from ``expanded"
parameterizations (\ref{eq:Fit-2}) -- (\ref{eq:Fit-3}) at
$\sqrt{s} \sim 3$ GeV than $B$. One can see the functions
(\ref{eq:Fit-2}) and (\ref{eq:Fit-3}) predict very close values of
$b$ in high-energy domain. Function (\ref{eq:Fit-4}) shows a
decreasing of $b$ at high energies in the contrast with $B$
parameter. In general estimations obtained with help of all
fitting functions are agree within errors up to the $\sqrt{s}=28$
TeV. But the large errors for function (\ref{eq:Fit-4}) do not
allow the unambiguous physics conclusion especially at the LHC
energies and above.
\subsection{$\Delta B$ and $NN$ data analysis}
\begin{figure*}
\begin{center}
\begin{tabular}{cc}
\mbox{
\resizebox{0.5\textwidth}{!}{%
  \includegraphics[width=8.0cm,height=8.0cm]{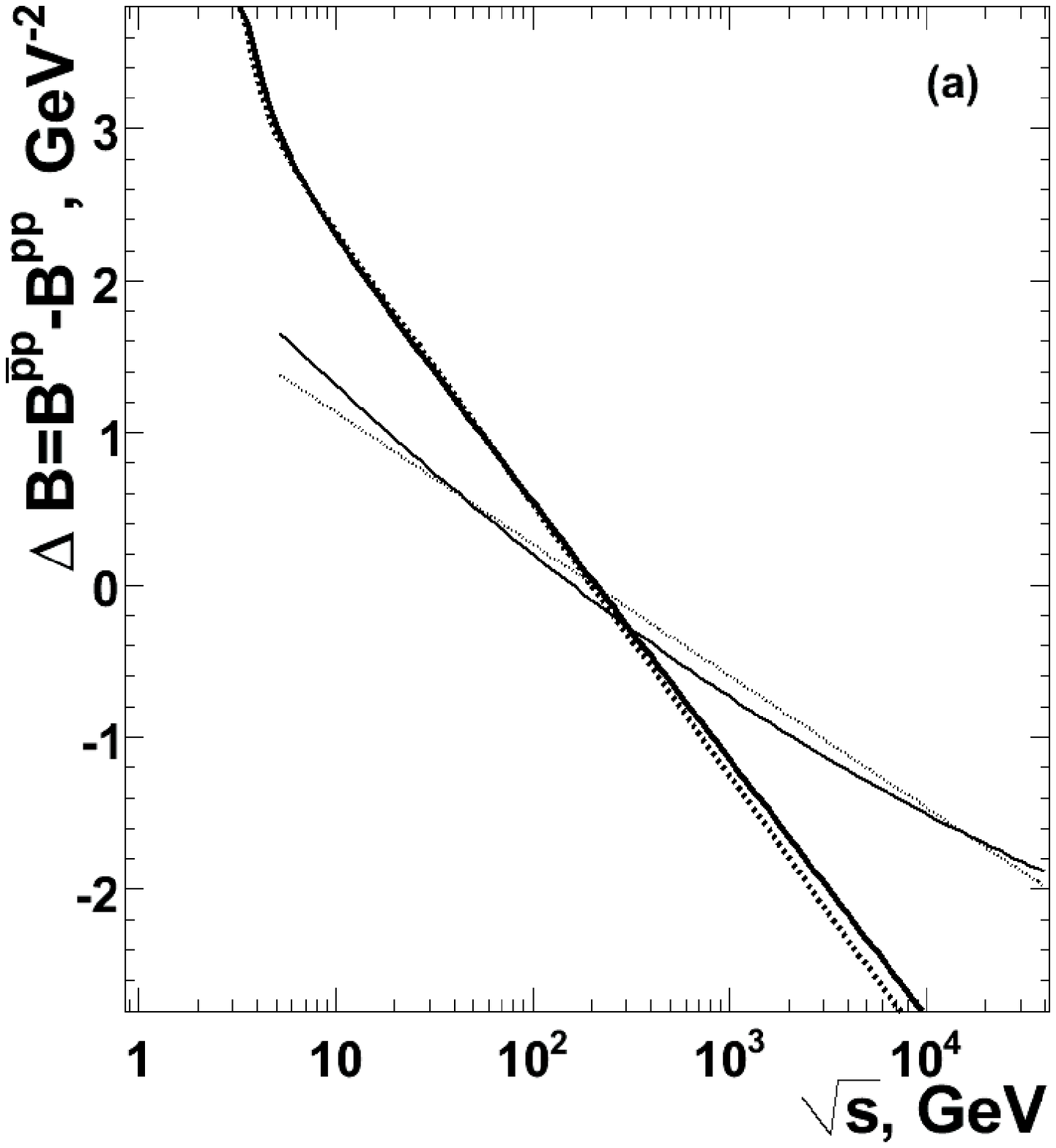}
}} & \mbox{
\resizebox{0.5\textwidth}{!}{%
  \includegraphics[width=8.0cm,height=8.0cm]{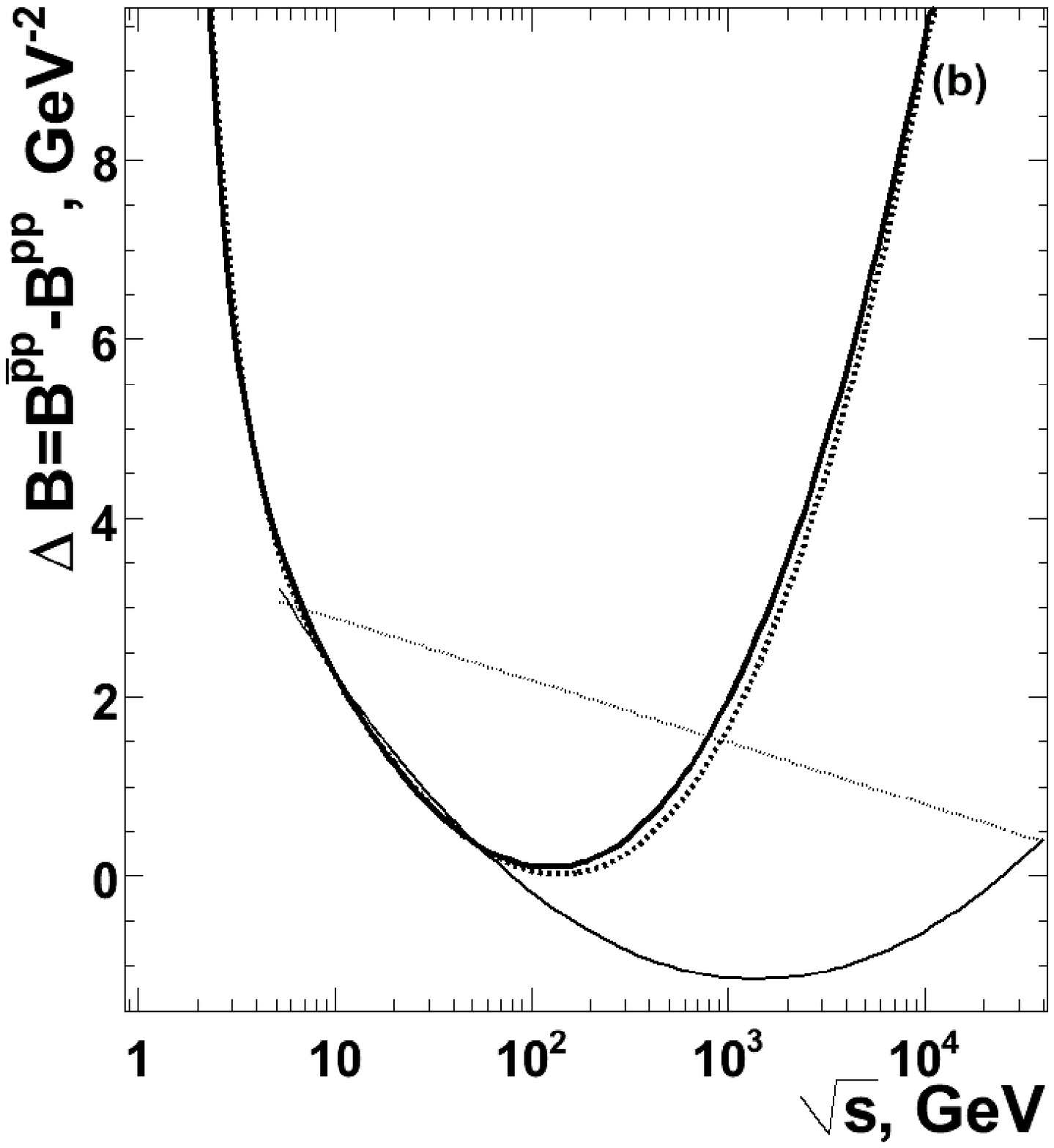}
}}
\end{tabular}
\end{center}
\vspace*{-0.4cm} \caption{The energy dependence of the difference
of elastic slopes for antiproton-proton and proton-proton
scattering in low $|t|$ domain (a) and in intermediate $|t|$ range
for exponential fit with index linear in $|t|$ of differential
cross-section (b). The correspondence of curves to the fit
functions is the same as well as in Figure 1.}\label{fig:6}
\end{figure*}

In accordance with rules from \cite{Okorokov-arXiv-0907.0951} the
difference of slopes $(\Delta B)$ for antiproton-proton and
proton-proton elastic scattering is calculated for each function
(\ref{eq:Fit-1}) -- (\ref{eq:Fit-4}) under study with parameters
from corresponded $\bar{p}p$ and $pp$ fits: $\Delta
B_{i}(s)=B^{\bar{p}p}_{i}(s)-B^{pp}_{i}(s),~i=\mbox{(2a) --
(2d)}$. The energy dependence of $\Delta B$ is shown at
Fig.\ref{fig:6}a and Fig.\ref{fig:6}b for low and intermediate
$|t|$ respectively. One can see that the difference of slopes
decreasing with increasing of energy for low $|t|$ domain
(Fig.\ref{fig:6}a) as well as in the previous analysis
\cite{Okorokov-arXiv-0907.0951}. The fitting functions
(\ref{eq:Fit-2}), (\ref{eq:Fit-3}) demonstrate much faster
decreasing of $\Delta B$ with increasing of $\sqrt{s}$ than that
the functions (\ref{eq:Fit-1}) and (\ref{eq:Fit-4}). At present
the proton-proton experimental data at highest available energy 8
TeV don't contradict with fast (square of logarithm of energy)
increasing of slope at high energies in general case. Such
behavior could be agreed with the asymptotic growth of total cross
section. Furthermore in contrast with the previous analysis
\cite{Okorokov-arXiv-0907.0951}, here the quadratic in $\ln
\varepsilon$ function (\ref{eq:Fit-4}) leads to much smaller
difference $\Delta B$ for $\bar{p}p$ and $pp$ scattering in high
energy domain for both low (Fig.\ref{fig:6}a) and intermediate
(Fig.\ref{fig:6}b) values of $|t|$. The Pomeron inspired function
(\ref{eq:Fit-1}) only predicts the decreasing of $\Delta B$ with
energy growth at intermediate $|t|$ (Fig.\ref{fig:6}b) for any
values of $\sqrt{s}$. The parameterizations (\ref{eq:Fit-2}) --
(\ref{eq:Fit-4}) predict the decreasing of difference of slopes at
low and intermediate energies and fast increasing of $\Delta B$ at
high energies for intermediate $|t|$ domain (Fig.\ref{fig:6}b). As
expected the most slow changing of $\Delta B$ is predicted by
Pomeron inspired function (\ref{eq:Fit-1}) at asymptotic energies.
All fitting functions with experimentally inspired parameters
don't predict the constant zero values of $\Delta B$ at high
energies. But it should be emphasized that only separate fits were
made for experimental data for $pp$ and $\bar{p}p$ elastic
reactions above. These results indicate on the importance of
investigations at ultra-high energies both $pp$ and $\bar{p}p$
elastic scattering for study of many fundamental questions and
predictions related to the general asymptotic properties of
hadronic physics.

\begin{figure*}
\begin{center}
\begin{tabular}{cc}
\mbox{
\resizebox{0.5\textwidth}{!}{%
  \includegraphics[width=8.0cm,height=8.0cm]{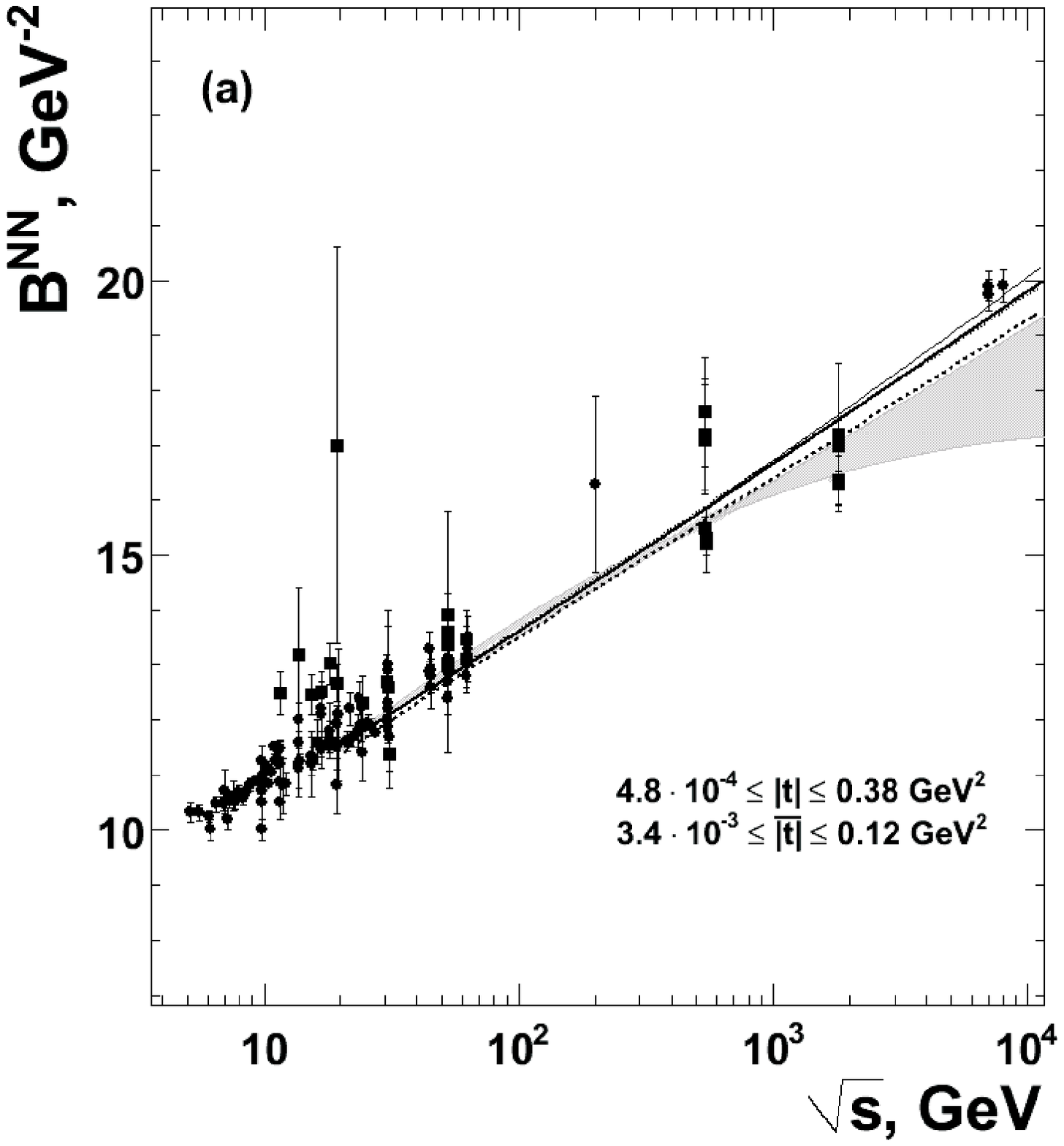}
}} & \mbox{
\resizebox{0.5\textwidth}{!}{%
  \includegraphics[width=8.0cm,height=8.0cm]{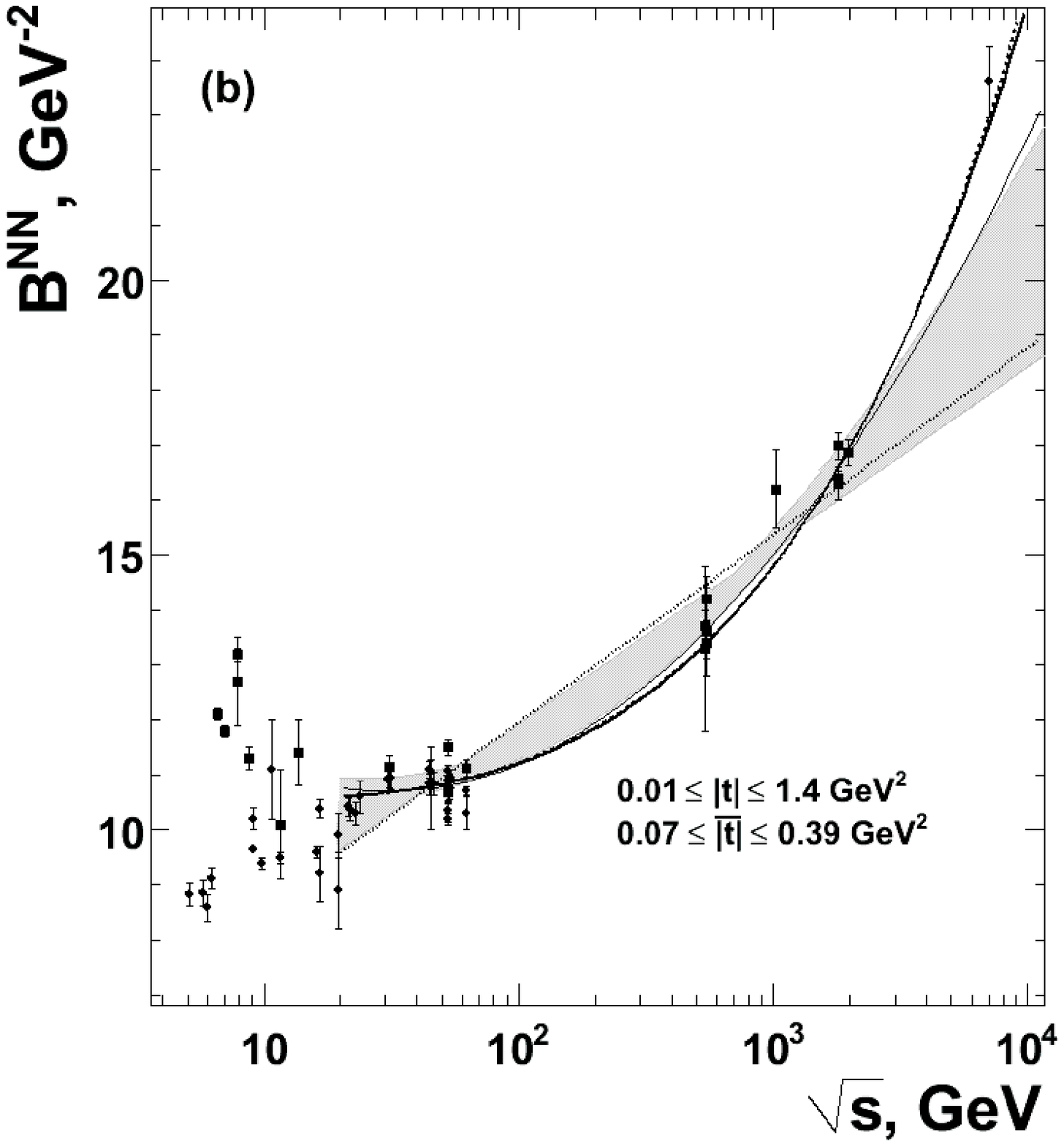}
}}
\end{tabular}
\end{center}
\vspace*{-0.4cm} \caption{Energy dependence of the elastic slope
parameter in nucleon-nucleon scattering for low (a) and
intermediate (b) domain of absolute values of momentum transfer.
Experimental points from fitted samples are indicated as
{\large$\bullet$} for $pp$ and as $\blacksquare$ for $\bar{p}p$.
Fits are shown for $\sqrt{s_{\mbox{\small{min}}}}=20$ GeV. The dot
curve is the fit of experimental slope by the function
(\ref{eq:Fit-1}), thick solid -- by the (\ref{eq:Fit-2}), dashed
-- by the (\ref{eq:Fit-3}), thin solid -- by the (\ref{eq:Fit-4}).
The shaded band corresponds to the spread of fitting functions for
previous analysis \cite{Okorokov-arXiv-0907.0951}.} \label{fig:7}
\end{figure*}

Also we have analyzed general data samples for $pp$ and $\bar{p}p$
elastic scattering. Slope parameters ($B$ and $b$) shows a
different energy dependence at $\sqrt{s} < 5$ GeV in proton-proton
and antiproton-proton elastic reactions in any $|t|$ domains under
study. Thus slopes for nucleon-nucleon data are investigated only
for $\sqrt{s} \geq 5$ GeV below. We have included in fitted
samples only $pp$ and $\bar{p}p$ points which have been included
in corresponding final data samples at separate study $pp$ and
$\bar{p}p$ elastic reactions above. We did not exclude any points
from $NN$ data sample, we change only the low energy boundary for
fitted domain. Experimental data for slope in nucleon-nucleon
elastic scattering against collision energy are shown in
Fig.\ref{fig:7}a at low $|t|$ and in Fig.\ref{fig:7}b for
intermediate $|t|$ together with fits by functions
(\ref{eq:Fit-1}) -- (\ref{eq:Fit-4}). We have fitted the general
nucleon-nucleon data sample at range of low energy boundary
$s_{\mbox{\small{min}}}=25, 100, 225$ and $400$ GeV$^{2}$. The fit
parameter values are indicated in Table \ref{tab:6} on the first
line for low boundary of the fitted energy domain
$s_{\mbox{\small{min}}}=25$ GeV$^{2}$ and on the second line --
for $s_{\mbox{\small{min}}}=400$ GeV$^{2}$. The fit quality
improves for most parameterizations under consideration at
increasing of $s_{\mbox{\small{min}}}$, thus fitting functions
(\ref{eq:Fit-1}) -- (\ref{eq:Fit-4}) are shown at Fig.\ref{fig:7}
for $s_{\mbox{\small{min}}}=400$ GeV$^{2}$.
As seen from Fig.\ref{fig:7}a there is no experimental data for
$\bar{p}p$ between $\sqrt{s}=5$ GeV and $\sqrt{s}=10$ GeV at low
$|t|$. This energy domain will available for further FAIR
facility. One need to emphasize the fit quality is some poorer
$\left(\chi^{2}/\mbox{n.d.f.} \! \simeq \! 2.3-2.9\right)$ at
$\sqrt{s} \! \geq \! 10$ GeV than that for $\sqrt{s} \! \geq \! 5$
GeV for functions (\ref{eq:Fit-1}) and (\ref{eq:Fit-3}). For all
cases of $s_{\mbox{\small{min}}}$ indicated above the value of the
$a_{1}$ parameter obtained with the function (\ref{eq:Fit-1})
agrees qualitatively with the prediction within the framework of
Pomeron model, but the value of the asymptotic shrinkage parameter
$(2\alpha'_{\cal{P}}=0.662 \pm 0.010)$ obtained at
$s_{\mbox{\small{min}}}=400$ GeV$^{2}$ is some larger than the
prediction for $\alpha'_{\cal{P}}$ from Pomeron inspired model for
TeV-energy domain \cite{Schegelsky-PRD-85-094024-2012}. Also
results from fit by function (\ref{eq:Fit-4}) with acceptable
quality at $s_{\mbox{\small{min}}}=400$ GeV$^{2}$ allow the
estimation $\left.2\alpha'_{\cal{P}}(s)\right|_{\sqrt{s}=8
\scriptsize{\mbox{TeV}}}=0.74 \pm 0.12$ which is some larger than
the corresponding prediction from
\cite{Schegelsky-PRD-85-094024-2012}. Furthermore the estimations
for $\left.2\alpha'_{\cal{P}}(s)\right|_{\sqrt{s}=8
\scriptsize{\mbox{TeV}}}$ do not depend on
$s_{\mbox{\small{min}}}$ within error bars.
At low $|t|$ all functions (\ref{eq:Fit-1}) -- (\ref{eq:Fit-4})
are close to each other at energies up to $\sqrt{s} \! \sim \! 1$
TeV at least and shows quasi-linear behavior for parameter values
obtained by fits with $s_{\mbox{\small{min}}}=25$ GeV$^{2}$ and
$s_{\mbox{\small{min}}}=400$ GeV$^{2}$. This observation confirms
the suggestion that $B^{NN} \propto \ln \varepsilon$ at high
$\sqrt{s}$ at low $|t|$ values. As seen from comparison between
the present fits and the spread of previous fit functions (shaded
band) there is a dramatic change of behavior of the fitting
function (\ref{eq:Fit-4}) in comparison with previous analysis
\cite{Okorokov-arXiv-0907.0951} due to experimental points at the
LHC energies. At present the fitting function (\ref{eq:Fit-4})
predicts increasing of the nuclear slope in high energy domain as
well as all other fitting functions under study. Such behavior is
opposite to the result of fit by function (\ref{eq:Fit-4}) of
experimental data sample at $\sqrt{s} \leq 1.8$ TeV
\cite{Okorokov-arXiv-0907.0951}. We have analyzed the
nucleon-nucleon data for slope parameter $B$ at intermediate $|t|$
values for exponential parametrization with index linear in $|t|$
of $d\sigma / dt$ (Fig.\ref{fig:7}b). As seen experimental $pp$
and $\bar{p}p$ data for $B$ differ significantly up to $\sqrt{s}
\simeq 10$ GeV at least that results in unacceptable fit qualities
for all functions under study ($\chi^{2}/\mbox{n.d.f.} \simeq
28.9$ for best fit by quadratic in logarithm function). The
Pomeron inspired function (\ref{eq:Fit-1}) contradicts with
experimental data at any $s_{\mbox{\small{min}}}$. We would like
to emphasize that the best fit quality for (\ref{eq:Fit-1}) is
obtained at $s_{\mbox{\small{min}}}=100$ GeV$^{2}$
$\left(\chi^{2}/\mbox{n.d.f.} \! \simeq \! 9.45\right)$ but it is
statistically unacceptable too. Functions (\ref{eq:Fit-2}) --
(\ref{eq:Fit-4}) agree with experimental dependence $B(s)$
reasonably and have a close fit qualities. Furthermore the
functions (\ref{eq:Fit-2}) and (\ref{eq:Fit-3}) demonstrate very
close behaviors in all range of $\sqrt{s}$ under consideration.
Best fit is ''expanded" function (\ref{eq:Fit-3}) in difference
with previous analysis \cite{Okorokov-arXiv-0907.0951}. As seen
from Fig.\ref{fig:7}b the slope parameter $B$ increases much
faster that $\ln \varepsilon$ at intermediate $|t|$ values. This
result confirms the suggestion made on the base of the
Fig.\ref{fig:3} above and allows us to derive the asymptotic
relation $B^{NN} \propto \ln^{k} \varepsilon, k > 2$ at high
$\sqrt{s}$ for intermediate $|t|$ domain. Fig.\ref{fig:7}b shows
that the experimental point at the LHC energy leads to faster
increasing of most of fitting functions in multi-TeV region in
comparison with previous analysis \cite{Okorokov-arXiv-0907.0951}.
Comparison between Fig.\ref{fig:7}a and Fig.\ref{fig:7}b
demonstrates that the domain of intermediate $|t|$ is preferable
with respect to the low $|t|$ range for discrimination of various
phenomenological parameterizations for $B(s)$, in particular, the
Pomeron inspired function (\ref{eq:Fit-1}) from ''expanded" ones.
At $s_{\mbox{\small{min}}}=400$ GeV$^{2}$ and intermediate $|t|$
values the following estimations of asymptotic shrinkage parameter
are obtained for fitting functions (\ref{eq:Fit-1}) and
(\ref{eq:Fit-4}) respectively: $2\alpha'_{\cal{P}}=0.706 \pm
0.012$ and $\left.2\alpha'_{\cal{P}}(s)\right|_{\sqrt{s}=7
\scriptsize{\mbox{TeV}}}=1.9 \pm 0.2$. As well as for low $|t|$
domain the values of $\alpha'_{\cal{P}}$ are equal within errors
at $\sqrt{s}=7$ TeV and $\sqrt{s}=8$ TeV in the case of the
function (\ref{eq:Fit-4}); the estimations for
$\left.2\alpha'_{\cal{P}}(s)\right|_{\sqrt{s}=7
\scriptsize{\mbox{TeV}}}$ do not depend on
$s_{\mbox{\small{min}}}$ within error bars.
The growth of $\alpha'_{\cal{P}}$ is observed with increasing of
$|t|~(|\bar{t}|)$ values, especially, for fitting function
(\ref{eq:Fit-4}).

\begin{table*}
\caption{Values of parameters for fitting of slope energy
dependence in $NN$ elastic scattering} \label{tab:6}
\begin{center}
\begin{tabular}{lccccc}
\hline \multicolumn{1}{l}{Function} &
\multicolumn{5}{c}{Parameter} \\
\cline{2-6} \rule{0pt}{10pt}
 & $B_{0}$, GeV$^{-2}$ & $a_{1}$, GeV$^{-2}$ & $a_{2}$, GeV$^{-2}$ & $a_{3}$ & $\chi^{2}/\mbox{n.d.f.}$ \\
\hline
\multicolumn{6}{c}{low $|t|$ domain} \\
\hline
(\ref{eq:Fit-1}) & $8.10 \pm 0.05$ & $0.301 \pm 0.004$ & --                & --               & $332/127$ \\
                 & $7.55 \pm 0.05$ & $0.331 \pm 0.005$ & --                & --               & $103/63$ \\
(\ref{eq:Fit-2}) & $9.4 \pm 0.2$   & $6.1 \pm 1.1$     & $-12 \pm 2$       & $-3.5 \pm 1.5$   & $281/125$ \\
                 & $8.1 \pm 0.7$   & $2.532 \pm 0.006$ & $-4.6 \pm 0.2$    & $0.988 \pm 0.015$& $103/61$ \\
(\ref{eq:Fit-3}) & $8.15 \pm 0.09$ & $0.208 \pm 0.010$ & $0.50 \pm 0.09$   & $0.118 \pm 0.005$& $281/125$ \\
                 & $7.8 \pm 0.2$   & $0.323 \pm 0.009$ & $-7.0 \pm 1.7$    & $-0.65 \pm 0.15$ & $102/61$ \\
(\ref{eq:Fit-4}) & $8.81 \pm 0.12$ & $0.204 \pm 0.015$ & $0.011 \pm 0.002$ & --               & $286/126$ \\
                 & $8.0 \pm 0.3$   & $0.28 \pm 0.04$   & $0.005 \pm 0.003$ & --               & $101/62$ \\
\hline
\multicolumn{6}{c}{intermediate $|t|$ domain} \\
\hline
(\ref{eq:Fit-1}) & $8.25 \pm 0.06$ & $0.214 \pm 0.004$ & --                & --               & $3135/57$ \\
                 & $5.17 \pm 0.11$ & $0.369 \pm 0.006$ & --                & --               & $353/36$ \\
(\ref{eq:Fit-2}) & $12.12 \pm 0.10$& $-0.217 \pm 0.11$ & $0.0097 \pm 0.0009$& $2.60 \pm 0.04$  & $1615/55$ \\
                 & $10.7 \pm 0.2$  & $-0.019 \pm 0.011$& $(1.04 \pm 0.17) \times 10^{-4}$& $4.08 \pm 0.08$& $107/34$ \\
(\ref{eq:Fit-3}) & $4.43 \pm 0.17$ & $-0.602 \pm 0.013$& $7.72 \pm 0.16$   & $0.0911 \pm 0.0009$& $1613/55$ \\
                 & $10.5 \pm 0.2$  & $-0.165 \pm 0.019$& $0.66 \pm 0.06$   & $0.187 \pm 0.004$ & $106/34$ \\
(\ref{eq:Fit-4}) & $13.07 \pm 0.14$& $-0.438 \pm 0.017$& $0.0743 \pm 0.0019$& --               & $1617/56$ \\
                 & $14.9 \pm 0.6$  & $-0.61 \pm 0.06$  & $0.089 \pm 0.006$ & --                & $118/35$ \\
\hline
\end{tabular}
\end{center}
\end{table*}

One can conclude the slope parameters for $pp$ and $\bar{p}p$
elastic scattering show universal behavior at $\sqrt{s} \geq 20$
GeV and ``expanded" functions represent the energy dependence for
both the low and the intermediate $|t|$ ranges for this energy
domain. Thus quantitative analysis of slopes at different $|t|$
allows us to get the following estimation of low energy boundary:
$\sqrt{s} \simeq 20$ GeV for universality of elastic
nucleon-nucleon scattering. This estimation agrees qualitatively
with both the results for differential cross-sections of $pp$ and
$\bar{p}p$ elastic reactions based on the crossing-symmetry and
derivative relations \cite{Okorokov-SPIN-2007} and the results for
global scattering parameters \cite{Okorokov-IJMPA-25-5333-2010}.
\section{Conclusions}
\label{sec:3}
The present status of diffraction slope parameter for elastic $pp$
and $\bar{p}p$ scattering is analyzed over the full energy domain
as well as predictions for some facilities.
The ''expanded" parameterizations allow us to describe
experimental $B$ at all available energies in low $|t|$ domain for
$pp$ as well as for intermediate $|t|$ values for $pp$ and
$\bar{p}p$ quite reasonably. The similar situation is observed for
fits of data samples joined for elastic $NN$ scattering. Therefore
``expanded" functions can be used as a reliable fits for wide
range of momentum transfer at all energies. The new LHC data lead
to dramatic change of behavior of quadratic in logarithm function
and usually to better agreement between various fitting function
in comparison with the analysis of $pp$ data at $\sqrt{s} \leq
200$ GeV. At low values of $|t|$ the ''standard" approximation
differs from expanded ones and experimental data mostly in the low
energy region; at intermediate $|t|$ this function is unacceptable
for fitting of experimental $NN$ data at all. The intermediate
$|t|$ range is preferable with respect to the low $|t|$ values for
discrimination of various phenomenological parameterizations for
$B(s)$ dependence. Based on the nuclear slope the low energy
boundary for universality of elastic nucleon-nucleon scattering is
estimated as $\sqrt{s} \simeq 20$ GeV for both the low and the
intermediate $|t|$ values that is in agreement with the hypothesis
of a universal shrinkage of the hadronic diffraction cone at high
energies.
The difference of slopes for antiproton-proton and proton-proton
elastic scattering $(\Delta B)$ shows the opposite behaviors at
high energies for low and intermediate $|t|$ domains (decreasing /
increasing, respectively) for all fitting functions with the
exception of Pomeron inspired one. All underlying fitting
functions with experimentally inspired values of parameters don't
predict the zero value for $\Delta B$ at both the low and the
intermediate $|t|$ ranges at high energies. Slop analysis of
joined $NN$ data samples allows us to estimate the asymptotic
shrinkage parameter parameter for various domains of $|t|$. The
estimation of the $\alpha'_{\cal{P}}$ obtained with quadratic in
logarithm function for $NN$ data at $\sqrt{s}=8$ TeV for low $|t|$
is noticeably larger than the expectation from the Pomeron theory.
But the growth of $\alpha'_{\cal{P}}$ with increasing of
$\sqrt{s}$ is observed from the comparison of the fit results from
present study and our earlier analysis for $\sqrt{s} \leq 1.8$
TeV. The such behavior of $\alpha'_{\cal{P}}$ agrees with Pomeron
inspired model. The growth of $\alpha'_{\cal{P}}$ is observed with
increasing of momentum transfer.
Based on the fit results the predictions for slope parameters $B$
and $b$ are obtained for elastic $pp$ and $\bar{p}p$ scattering in
energy domains of some facilities. It seems the phenomenological
model with hadronic amplitude corresponding to the exchange of
three pomerons describes the $B$ some closer to the experimental
fit inspired values at the LHC energy both at low and intermediate
$|t|$ than other models.

\section*{\hspace*{7.0cm}Appendix}
\vspace*{-0.2cm}
\section*{Experimental database for slope in nucleon-nucleon elastic scattering}\label{sec:4}

\hspace*{0.5cm}In this Appendix, the numerical values used in the
Sec.~\ref{sec:1} are shown for diffractive slope parameter. The
da\-ta\-ba\-se includes the experimental values of the slope
obtained with exponential approximations of $d\sigma /dt$ shown
above. In the Tables \ref{tab:app1} -- \ref{tab:app4} the
statistical errors are shown by first and available systematic
uncertainties are demonstrated by second. The compilation of some
early experimental measurements of the slope in $pp / \bar{p}p$
elastic scattering at low energies was made in
\cite{Lasinski-NPB-37-1-1972}. This one reference is shown and
detail list of corresponding original papers can be found in
\cite{Lasinski-NPB-37-1-1972}. In the case of $\bar{p}p$ elastic
scattering in low $|t|$ domain (Table \ref{tab:app3}) the average
$\sqrt{s}$ are demonstrated for ranges of initial momentum
($p_{\,\scriptsize{\mbox{lab}}}$) 1.23 -- 1.33 GeV/$c$ and 1.45 --
1.65 GeV/$c$ \cite{Lasinski-NPB-37-1-1972}. As for database of
collective parameters \cite{Okorokov-IJMPA-27-1250037-2012} the
results from latest paper of the experiment are included in Tables
\ref{tab:app1} -- \ref{tab:app4} below at certain $\sqrt{s}$ and
range of $|t|$. Measurements of particular experiment can be
separated among Tables \ref{tab:app1} -- \ref{tab:app4} in
accordance with the limits for low / intermediate $|t|$ domain
defined above. But here there is no averaging over results of
various experiments at equal collision energy in difference with
\cite{Okorokov-IJMPA-27-1250037-2012}. Furthermore the averaging
over results in different ranges of $|t|$ for particular
experiment at the same $\sqrt{s}$ is absent too within the low /
intermediate $|t|$ domain. Thus the measurements of particular
experiment at fixed $\sqrt{s}$ and in various ranges of $|t|$ are
shown separately in the corresponding Table below if these
measurements are in the chosen $|t|$ domain (see, for example,
\cite{Barish-PRD-9-1171-1974,Bozzo-PLB-147-385-1984,Amos-PLB-247-127-1990})
and agree with general trend, at least, qualitatively. Therefore
the database includes the differential measurements over both the
$\sqrt{s}$ and the $|t|$ in the cases of some experiments.

\begin{table*}[h!]
\caption{Experimental $B_{pp}$ for low $|t|$ domain}
\label{tab:app1}
\begin{center}
\begin{tabular}{llc|llc|llc}
\hline \multicolumn{1}{c}{$\sqrt{s}$, GeV} &
\multicolumn{1}{c}{$B_{pp}$, GeV$^{-2}$} &
\multicolumn{1}{c|}{Ref.} & \multicolumn{1}{c}{$\sqrt{s}$, GeV} &
\multicolumn{1}{c}{$B_{pp}$, GeV$^{-2}$} &
\multicolumn{1}{c|}{Ref.} & \multicolumn{1}{c}{$\sqrt{s}$, GeV} &
\multicolumn{1}{c}{$B_{pp}$, GeV$^{-2}$} &
\multicolumn{1}{c}{Ref.} \rule{0pt}{10pt}\\
\hline
1.962 & $1.60 \pm 0.60$ & \cite{Lasinski-NPB-37-1-1972} & 6.977 & $10.70 \pm 0.40$ & \cite{Bartenev-PRL-29-1755-1972} & 19.42 & $10.80 \pm 0.50$ & \cite{Bartenev-PRL-29-1755-1972}\rule{0pt}{10pt}\\
1.974 & $1.43 \pm 0.41$ & \cite{Leihth-PreprintSLAC-SLAC-PUB-1263-1973}& 7.167 & $10.20 \pm 0.20$ & \cite{Bellettini-PL-14-164-1965}& 19.42 & $11.50 \pm 0.40$ & \cite{Cool-PRD-24-2821-1981}\\
1.993 & $0.40 \pm 0.50$ & \cite{Lasinski-NPB-37-1-1972}& 7.311 & $10.52 \pm 0.12 \pm 0.30$ & \cite{Beznogikh-NPB-54-78-1973}& 19.46 & $11.93 \pm 0.05 \pm 0.32$ & \cite{Fajardo-PRD-24-46-1981} \\
2.000 & $0.38 \pm 0.19$ & \cite{Leihth-PreprintSLAC-SLAC-PUB-1263-1973}& 7.620 & $10.61 \pm 0.27$ & \cite{Geshkov-PRD-13-1846-1976}& 19.66 & $12.10 \pm 1.20$ & \cite{Barish-PRD-9-1171-1974}\\
2.002 & $0.60 \pm 0.20$ & \cite{Lasinski-NPB-37-1-1972}& 7.664 & $10.49 \pm 0.12 \pm 0.30$ & \cite{Beznogikh-NPB-54-78-1973}& 21.22 & $11.61 \pm 0.19 \pm 0.20$ & \cite{Bartenev-PRL-31-1088-1973} \\
2.069 & $0.99 \pm 0.38$ & \cite{Leihth-PreprintSLAC-SLAC-PUB-1263-1973}& 8.019 & $10.69 \pm 0.12 \pm 0.30$& & 21.49 & $11.57 \pm 0.03$ & \cite{Barbiellini-PLB-39-663-1972}\\
2.069 & $1.46 \pm 0.47$ & & 8.345 & $10.57 \pm 0.11 \pm 0.30$ & & 21.70 & $12.20 \pm 0.30$ & \cite{Burq-PLB-109-124-1982}\\
2.083 & $0.64 \pm 0.29$ & & 8.550 & $10.68 \pm 0.09 \pm 0.30$ & & 22.55 & $11.69 \pm 0.10 \pm 0.20$ & \cite{Bartenev-PRL-31-1088-1973}\\
2.125 & $2.63 \pm 0.37$ & & 8.832 & $10.82 \pm 0.11 \pm 0.30$ & & 23.50 & $11.80 \pm 0.30$ & \cite{Amos-NPB-262-689-1985}\\
2.177 & $3.56 \pm 1.09$ & \cite{Velichko-JEPT-33-598-1981} & 9.030$^{*}$ & $13.50 \pm 0.74$ & \cite{Apokin-YaF-25-94-1977}& 23.60 & $11.80 \pm 0.30$ & \cite{Amaldi-PLB-62-460-1976}\\
2.218 & $4.42 \pm 0.51$ & & 9.303 & $10.90 \pm 0.09 \pm 0.30$ & \cite{Beznogikh-NPB-54-78-1973}& 23.76 & $12.40 \pm 0.30$ & \cite{Burq-PLB-109-124-1982}\\
2.239 & $5.83 \pm 0.39$ & & 9.777 & $10.50 \pm 0.40$ & \cite{Bartenev-PRL-29-1755-1972}& 24.23 & $11.90 \pm 0.28 \pm 0.20$ & \cite{Bartenev-PRL-31-1088-1973}\\
2.259 & $4.88 \pm 0.45$ & & 9.777 & $11.25 \pm 0.28$ & \cite{Geshkov-PRD-13-1846-1976}& 24.30 & $11.40 \pm 0.50 \pm 0.07$ & \cite{Breedon-PLB-216-459-1989}\\
2.279 & $5.53 \pm 0.46$ & & 9.778 & $10.70 \pm 0.18 \pm 0.20$ & \cite{Bartenev-PRL-31-1088-1973}& 25.59 & $11.96 \pm 0.15 \pm 0.20$ & \cite{Bartenev-PRL-31-1088-1973}\\
2.300 & $6.49 \pm 0.60$ & & 9.778 & $10.00 \pm 0.20 \pm 0.15$ & \cite{Ayres-PRD-15-3105-1977}& 26.42 & $11.87 \pm 0.15 \pm 0.20$ & \\
2.319 & $6.24 \pm 0.37$ & & 9.837 & $10.84 \pm 0.11 \pm 0.30$ & \cite{Beznogikh-NPB-54-78-1973}& 27.29 & $11.77 \pm 0.10 \pm 0.20$ & \\
2.465 & $5.70 \pm 0.60$ & \cite{Lasinski-NPB-37-1-1972}& 9.977 & $11.00 \pm 0.12 \pm 0.30$ & & 30.48 & $12.00 \pm 0.20$ & \cite{Holder-PLB-36-400-1971}\\
2.465 & $5.97 \pm 0.15$ & \cite{Zhurkin-YaF-28-1280-1978}& 9.987$^{*}$ & $12.70 \pm 0.70$ & \cite{Apokin-YaF-25-94-1977}& 30.60 & $12.00 \pm 0.20$ & \cite{Baksay-NPB-141-1-1978}\\
2.696 & $7.80 \pm 0.50$ & \cite{Lasinski-NPB-37-1-1972}& 10.19 & $11.12 \pm 0.13 \pm 0.30$ & \cite{Beznogikh-NPB-54-78-1973}& 30.60 & $12.20 \pm 0.30$ & \cite{Amos-NPB-262-689-1985}\\
2.702 & $7.60 \pm 0.43$ & \cite{Beznogikh-PLB-43-85-1973} & 10.43 & $11.11 \pm 0.10 \pm 0.30$ & & 30.70 & $12.90 \pm 1.10$ & \cite{Favart-PRL-47-1191-1981}\\
2.725 & $7.80 \pm 0.37$ & \cite{Leihth-PreprintSLAC-SLAC-PUB-1263-1973}& 10.52 & $10.83 \pm 0.07 \pm 0.20$ & \cite{Bartenev-PRL-31-1088-1973} & 30.80 & $12.30 \pm 0.30$ & \cite{Amaldi-PLB-62-460-1976}\\
2.735 & $6.50 \pm 0.60$ & \cite{Lasinski-NPB-37-1-1972}& 10.71 & $11.05 \pm 0.08 \pm 0.30$ & \cite{Beznogikh-NPB-54-78-1973}& 30.81 & $11.87 \pm 0.28$ & \cite{Barbiellini-PLB-39-663-1972}\\
2.987 & $8.33 \pm 0.37$ & & 11.00 & $11.50 \pm 0.11 \pm 0.30$ &\cite{Beznogikh-PLB-43-85-1973}& 30.82 & $13.00 \pm 0.70$ & \cite{Amaldi-PLB-36-504-1971}  \\
3.266 & $7.69 \pm 0.37$ & \cite{Leihth-PreprintSLAC-SLAC-PUB-1263-1973}& 11.17 & $11.50 \pm 0.11 \pm 0.30$ &\cite{Beznogikh-NPB-54-78-1973} & 31.00 & $11.70 \pm 0.62$ & \cite{Breakstone-NPB-248-253-1984}\\
3.307 & $7.80 \pm 0.44$ & \cite{Beznogikh-PLB-43-85-1973}& 11.22 & $11.24 \pm 0.11 \pm 0.30$ & & 44.70 & $12.80 \pm 0.30$ & \cite{Amaldi-PLB-66-390-1977}\\
3.321 & $7.70 \pm 0.50$ & \cite{Lasinski-NPB-37-1-1972}& 11.47 & $11.46 \pm 0.09 \pm 0.30$ & & 44.87 & $12.87 \pm 0.20$ & \cite{Barbiellini-PLB-39-663-1972}\\
3.843 & $9.60 \pm 0.30$ & & 11.53 & $11.48 \pm 0.15 \pm 0.30$ & & 44.90 & $13.30 \pm 0.30$ & \cite{Baksay-NPB-141-1-1978}\\
3.851 & $9.14 \pm 0.35$ & \cite{Beznogikh-PLB-43-85-1973}& 11.54 & $11.21 \pm 0.40$ & \cite{Geshkov-PRD-13-1846-1976}& 45.00 & $12.60 \pm 0.40$ & \cite{Amaldi-PLB-44-112-1973}\\
3.874$^{*}$ & $8.20 \pm 0.20$ & \cite{Lasinski-NPB-37-1-1972}& 11.54 & $10.50 \pm 0.30 \pm 0.16$ & \cite{Ayres-PRD-15-3105-1977}& 45.00 & $12.80 \pm 0.30$ & \cite{Amaldi-PLB-62-460-1976}\\
3.875 & $9.40 \pm 0.60$ & \cite{Jenni-NPB-129-232-1977}& 11.54 & $10.87 \pm 0.14 \pm 0.34$ & \cite{Fajardo-PRD-24-46-1981}& 45.06 & $12.90 \pm 0.40$ & \cite{Amaldi-PLB-36-504-1971}\\
3.995 & $9.53 \pm 0.23$ & \cite{Leihth-PreprintSLAC-SLAC-PUB-1263-1973}& 11.94 & $10.80 \pm 0.50$ & \cite{Bartenev-PRL-29-1755-1972}& 52.80 & $13.10 \pm 0.20$ & \cite{Baksay-NPB-141-1-1978}\\
4.305 & $9.52 \pm 0.26$ & \cite{Lasinski-NPB-37-1-1972}& 12.17 & $10.84 \pm 0.20 \pm 0.20$ & \cite{Bartenev-PRL-31-1088-1973}& 52.80 & $13.10 \pm 0.20$ & \cite{Amaldi-NPB-166-301-1980}\\
4.307 & $9.40 \pm 0.30$ & \cite{Beznogikh-PLB-43-85-1973}& 13.76 & $11.60 \pm 0.50$ & \cite{Bartenev-PRL-29-1755-1972}& 52.80 & $12.70 \pm 0.20$ & \cite{Favart-PRL-47-1191-1981}\\
4.329 & $8.72 \pm 0.38 \pm 0.20$ & \cite{Bartenev-PRL-31-1088-1973}& 13.76 & $11.20 \pm 0.60$ & \cite{Cool-PRD-24-2821-1981}& 52.80 & $13.09 \pm 0.37 \pm 0.21$ & \cite{Ambrosio-PLB-115-495-1982}\\
4.478 & $9.50 \pm 0.22$ & \cite{Leihth-PreprintSLAC-SLAC-PUB-1263-1973}& 13.76 & $11.13 \pm 0.14 \pm 0.33$ & \cite{Fajardo-PRD-24-46-1981}& 52.80 & $12.87 \pm 0.14$ & \cite{Amos-NPB-262-689-1985} \\
4.540 & $9.20 \pm 0.40$ & \cite{Jenni-NPB-129-232-1977}& 13.76 & $12.00 \pm 0.30$ & \cite{Burq-NPB-217-285-1983}&  52.90 & $13.10 \pm 0.30$ & \cite{Amaldi-PLB-66-390-1977}\\
4.562 & $10.40 \pm 0.40$ & \cite{Bellettini-PL-14-164-1965}& 13.90 & $11.24 \pm 0.13 \pm 0.20$ & \cite{Bartenev-PRL-31-1088-1973}&  52.99 & $12.40 \pm 0.30$ & \cite{Barbiellini-PLB-39-663-1972}\\
4.621 & $8.90 \pm 0.17 \pm 0.10$ & \cite{Carnegie-PLB-59-313-1975}& 15.36 & $11.34 \pm 0.09 \pm 0.33$ & \cite{Fajardo-PRD-24-46-1981}& 53.00 & $13.10 \pm 0.30$ & \cite{Amaldi-PLB-44-112-1973} \\
4.721 & $9.25 \pm 0.45$ & \cite{Lasinski-NPB-37-1-1972}& 15.37 & $11.20 \pm 0.60$ & \cite{Bartenev-PRL-29-1755-1972}& 53.10 & $13.00 \pm 0.30$ & \cite{Amaldi-PLB-36-504-1971}\\
4.721 & $9.16 \pm 0.37$ & \cite{Beznogikh-PLB-43-85-1973}& 15.56 & $11.30 \pm 0.20 \pm 0.20$ & \cite{Bartenev-PRL-31-1088-1973}& 62.30 & $13.02 \pm 0.27$ & \cite{Amos-NPB-262-689-1985}\\
4.934 & $9.03 \pm 0.30 \pm 0.20$ & \cite{Bartenev-PRL-31-1088-1973}& 16.26 & $11.60 \pm 0.40 \pm 0.17$ & \cite{Ayres-PRD-15-3105-1977}& 62.40 & $13.30 \pm 0.30$ & \cite{Amaldi-PLB-66-390-1977} \\
4.953 & $10.80 \pm 1.00$ & \cite{Lasinski-NPB-37-1-1972}& 16.83 & $12.20 \pm 0.50$ & \cite{Bartenev-PRL-29-1755-1972}& 62.50 & $12.80 \pm 0.20$ & \cite{Baksay-NPB-141-1-1978}\\
4.953 & $9.24 \pm 0.35$ & \cite{Leihth-PreprintSLAC-SLAC-PUB-1263-1973}& 16.83 & $11.57 \pm 0.23 \pm 0.20$ & \cite{Bartenev-PRL-31-1088-1973}& 63.00 & $13.30 \pm 0.60$ & \cite{Ambrosio-PLB-113-347-1982}\\
5.150 & $10.32 \pm 0.17 \pm 0.30$ & \cite{Beznogikh-NPB-54-78-1973}& 16.83 & $12.10 \pm 0.30$ & \cite{Burq-NPB-217-285-1983}& 200.0 & $16.30 \pm 1.60 \pm 0.90$ & \cite{Bultmann-PLB-579-245-2004}\\
5.562 & $10.31 \pm 0.15 \pm 0.30$ & & 16.91 & $11.46 \pm 0.06 \pm 0.33$ & \cite{Fajardo-PRD-24-46-1981}&  7000  & $19.89 \pm 0.02 \pm 0.27$ & \cite{Antchev-LJEFP-101-21002-2013}\\
6.105 & $10.24 \pm 0.11 \pm 0.30$ & & 18.14 & $11.63 \pm 0.09 \pm 0.33$ & &  7000  & $19.73 \pm 0.14 \pm 0.26$ & \cite{Aad-NPB-889-486-2014}\\
6.171 & $10.00 \pm 0.20$ & \cite{Bellettini-PL-14-164-1965}& 18.17 & $11.80 \pm 0.50$ & \cite{Bartenev-PRL-29-1755-1972}& 8000 & $19.90 \pm 0.30$ & \cite{Antchev-PRL-111-021001-2013}\\
6.522 & $10.47 \pm 0.14 \pm 0.30$ & \cite{Beznogikh-NPB-54-78-1973}& 18.17 & $11.52 \pm 0.11 \pm 0.20$ & \cite{Bartenev-PRL-31-1088-1973} & & & \\
6.920 & $10.48 \pm 0.13 \pm 0.30$ & & 19.37 & $11.56 \pm 0.12 \pm 0.20$ & & & & \\
\hline \multicolumn{9}{l}{$^*$\rule{0pt}{11pt}\footnotesize The
point is excluded from the fit procedure.}
\end{tabular}
\end{center}
\end{table*}

\begin{table*}[h!]
\caption{Experimental $B_{pp}$ for intermediate $|t|$ domain}
\label{tab:app2}
\begin{center}
\begin{tabular}{llc|llc|llc}
\hline \multicolumn{1}{c}{$\sqrt{s}$, GeV} &
\multicolumn{1}{c}{$B_{pp}$, GeV$^{-2}$} &
\multicolumn{1}{c|}{Ref.} & \multicolumn{1}{c}{$\sqrt{s}$, GeV} &
\multicolumn{1}{c}{$B_{pp}$, GeV$^{-2}$} &
\multicolumn{1}{c|}{Ref.} & \multicolumn{1}{c}{$\sqrt{s}$, GeV} &
\multicolumn{1}{c}{$B_{pp}$, GeV$^{-2}$} &
\multicolumn{1}{c}{Ref.} \rule{0pt}{10pt}\\
\hline
\multicolumn{9}{c}{Linear parameterization $\ln(d\sigma / dt) \propto (-B|t|)$} \rule{0pt}{10pt}\\
\hline
2.034 & $1.60 \pm 0.60$ & \cite{Lasinski-NPB-37-1-1972} & 3.942 & $7.37 \pm 0.29$ & & 23.00 & $10.30 \pm 0.20$ & ~\cite{Kwak-PLB-58-233-1975} \rule{0pt}{10pt}\\
2.037 & $0.20 \pm 1.40$ & & 4.220 & $7.75 \pm 0.11$ & \cite{Harting-NC-38-60-1965} & 23.76 & $10.60 \pm 0.30$ & \cite{Firestone-PRD-10-2080-1974} \\
2.058 & $0.59 \pm 0.11$ & & 4.331 & $8.39 \pm 20$ & \cite{Leihth-PreprintSLAC-SLAC-PUB-1263-1973}& 30.81 & $10.91 \pm 0.22$ & \cite{Barbiellini-PLB-39-663-1972} \\
2.066 & $1.00 \pm 0.50$ & & 4.519 & $8.95 \pm 0.06$ & \cite{Edelstein-PRD-5-1073-1972} & 31.00 & $10.92 \pm 0.15$ & \cite{Breakstone-NPB-248-253-1984} \\
2.075 & $0.00 \pm 0.70$ & & 4.621 & $8.11 \pm 0.17 \pm 0.10$ & \cite{Carnegie-PLB-59-313-1975} & 31.39 & $10.93 \pm 0.20$ & \cite{Leihth-PreprintSLAC-SLAC-PUB-1263-1973} \\
2.098 & $0.90 \pm 0.40$ & & 4.694 & $8.53 \pm 0.22$ & \cite{Leihth-PreprintSLAC-SLAC-PUB-1263-1973} & 44.61 & $11.10 \pm 0.15$ & \cite{Holder-PLB-36-400-1971}  \\
2.112 & $0.40 \pm 0.60$ & & 5.010$^{*}$ & $8.19 \pm 0.13$ & \cite{Harting-NC-38-60-1965} & 44.87 & $10.83 \pm 0.20$ & \cite{Barbiellini-PLB-39-663-1972} \\
2.132 & $1.50 \pm 0.60$ & & 5.122 & $8.83 \pm 20$ & \cite{Leihth-PreprintSLAC-SLAC-PUB-1263-1973} & 45.00 & $10.75 \pm 0.75$ & \cite{Holder-PLB-35-355-1971}\\
2.142 & $1.20 \pm 0.40$ & & 5.780 & $8.86 \pm 23$ & & 45.41 & $11.08 \pm 0.20$ & \cite{Leihth-PreprintSLAC-SLAC-PUB-1263-1973} \\
2.153 & $1.30 \pm 0.30$ & & 6.028 & $8.58 \pm 0.24$ & \cite{Harting-NC-38-60-1965} & 46.34 & $10.84 \pm 0.20$ & \\
2.166 & $3.00 \pm 0.50$ & & 6.223 & $9.12 \pm 0.19$ & \cite{Leihth-PreprintSLAC-SLAC-PUB-1263-1973} & 52.61 & $10.90 \pm 0.15$ & \cite{Holder-PLB-36-400-1971} \\
2.180 & $2.70 \pm 0.50$ & & 9.081 & $9.65 \pm 0.07$ & \cite{Nurushev-Proceedings-ICHEP-1974}& 52.80& $10.70 \pm 0.20$ & \cite{Amaldi-NPB-166-301-1980} \\
2.191 & $3.60 \pm 0.60$ & & 9.081 & $10.20 \pm 20$ & \cite{Diddens-Proceedings-ICHEP-1974}& 52.80& $10.34 \pm 0.19 \pm 0.06$ & \cite{Ambrosio-PLB-115-495-1982}\\
2.201$^{*}$ & $5.10 \pm 0.60$ & & 9.778 & $9.40 \pm 0.10 \pm 0.14$ & \cite{Ayres-PRD-15-3105-1977} & 52.99 & $10.80 \pm 0.20$ & \cite{Barbiellini-PLB-39-663-1972} \\
2.688 & $5.55 \pm 0.20$ & \cite{Leihth-PreprintSLAC-SLAC-PUB-1263-1973}& 10.69 & $11.10 \pm 0.90$ & \cite{Bromberg-PRD-15-64-1977}& 53.00 & $11.06 \pm 0.11$ & \cite{Breakstone-NPB-248-253-1984}\\
2.780 & $6.46 \pm 0.25$ & & 11.54 & $9.50 \pm 0.10 \pm $ 0.14& \cite{Ayres-PRD-15-3105-1977} & 53.00 & $10.20 \pm 0.10$ & \cite{DeKerret-PLB-68-374-1977} \\
2.837 & $5.55 \pm 0.23$ & & 16.26 & $9.60 \pm 0.10 \pm 0.14$ & & 53.35 & $10.78 \pm 0.23$ & \cite{Leihth-PreprintSLAC-SLAC-PUB-1263-1973}\\
3.034 & $6.25 \pm 0.26$ & & 16.66 & $10.38 \pm 0.17$ & & 54.44 & $10.93 \pm 0.18$ & \\
3.363 & $7.12 \pm 0.11$ & \cite{Lasinski-NPB-37-1-1972}& 16.66 & $9.20 \pm 0.50$ & \cite{Brick-PRD-25-2794-1982}& 62.00 & $10.30 \pm 0.30$ & \cite{Kwak-PLB-58-233-1975}\\
3.378 & $7.37 \pm 0.25$ & \cite{Leihth-PreprintSLAC-SLAC-PUB-1263-1973}& 19.66 & $8.90 \pm 0.70$ & \cite{Barish-PRD-9-1171-1974}& 62.00 & $10.71 \pm 0.08$ & \cite{Breakstone-NPB-248-253-1984}\\
3.433 & $7.60 \pm 0.50$ & \cite{Lasinski-NPB-37-1-1972}& 19.66 & $9.90 \pm 0.40$ & & 7000 & $23.60 \pm 0.50 \pm 0.40$ & \cite{Antchev-LJEFP-95-41001-2011}\\
3.778 & $7.94 \pm 0.26$ & \cite{Colton-PRD-7-3267-1973}& 21.49 & $10.42 \pm 0.17$ & \cite{Barbiellini-PLB-39-663-1972} & & & \\
3.916 & $7.95 \pm 0.18$ & \cite{Leihth-PreprintSLAC-SLAC-PUB-1263-1973}& 21.84 & $10.38 \pm 0.22$ & \cite{Leihth-PreprintSLAC-SLAC-PUB-1263-1973}& & & \\
\hline
\multicolumn{9}{c}{Quadratic parameterization $\ln(d\sigma / dt) \propto (-B|t| \pm Ct^{2})$} \rule{0pt}{10pt}\\
\hline
2.212 & $4.60 \pm 0.90$ & \cite{Lasinski-NPB-37-1-1972} & 3.826 & $9.78 \pm 0.21$ & \cite{Foley-PRL-11-425-1963}& 6.272 & $9.17 \pm 0.11$ & \cite{Edelstein-PRD-5-1073-1972} \rule{0pt}{10pt}\\
2.236 & $5.50 \pm 0.70$ & & 4.131 & $7.00 \pm 1.10$ & \cite{Lasinski-NPB-37-1-1972}& 6.434 & $9.50 \pm 0.90$ & \cite{Lasinski-NPB-37-1-1972} \\
2.265 & $6.40 \pm 0.60$ & & 4.220 & $8.16 \pm 0.28$ & \cite{Harting-NC-38-60-1965}& 6.434 & $9.80 \pm 0.30$ & \\
2.272 & $7.64 \pm 0.46$ & & 4.220 & $8.35 \pm 0.25$ & & 6.477 & $10.90 \pm 0.60$ & \\
2.304 & $6.50 \pm 0.60$ & & 4.286 & $9.62 \pm 0.22$ & \cite{Foley-PRL-11-425-1963}& 6.547 & $9.63 \pm 0.78$ & \cite{Foley-PRL-15-45-1965}\\
2.307 & $8.12 \pm 0.84$ & & 4.701 & $9.79 \pm 0.23$ & & 6.871 & $9.90 \pm 1.00$ & \cite{Lasinski-NPB-37-1-1972}\\
2.311 & $6.50 \pm 0.80$ & & 4.729 & $8.56 \pm 0.47$ & \cite{Foley-PRL-15-45-1965}& 6.929 & $7.97 \pm 1.56$ & \cite{Foley-PRL-15-45-1965}\\
2.325 & $7.70 \pm 0.30$ & & 5.010 & $9.05 \pm 0.34$ & \cite{Harting-NC-38-60-1965}& 7.138 & $11.60 \pm 0.70$ & \cite{Lasinski-NPB-37-1-1972} \\
2.339 & $6.30 \pm 0.50$ & & 5.010 & $9.71 \pm 0.16$ & & 7.422 & $9.10 \pm 1.10$ & \\
2.342 & $8.80 \pm 1.00$ & & 5.084 & $10.03 \pm 0.28$ & \cite{Foley-PRL-11-425-1963}& 7.584 & $9.34 \pm 0.28$ & \cite{Edelstein-PRD-5-1073-1972}\\
2.360 & $5.70 \pm 0.50$ & & 5.120 & $9.90 \pm 1.10$ & \cite{Lasinski-NPB-37-1-1972}& 9.081$^{*}$ & $11.20 \pm 0.22$ & \cite{Nurushev-Proceedings-ICHEP-1974}\\
2.395 & $6.30 \pm 0.40$ & & 5.128 & $11.40 \pm 1.20$ & & 9.778 & $10.30 \pm 0.10 \pm 0.15$ & \cite{Ayres-PRD-15-3105-1977}\\
2.458 & $7.10 \pm 0.40$ & & 5.440 & $10.37 \pm 0.33$ & \cite{Foley-PRL-11-425-1963}& 11.54 & $10.60 \pm 0.20 \pm 0.16$ & \\
2.510 & $7.30 \pm 1.00$ & & 5.462 & $8.89 \pm 0.52$ & \cite{Foley-PRL-15-45-1965}& 13.76 & $10.70 \pm 0.20 \pm 0.16$ & \\
2.682 & $8.10 \pm 0.40$ & & 5.491 & $8.81 \pm 0.25$ & \cite{Edelstein-PRD-5-1073-1972}& 13.90 & $11.40 \pm 0.70$ & \cite{Bromberg-PRL-31-1563-1973}\\
2.768 & $6.10 \pm 0.90$ & & 5.559 & $10.00 \pm 0.80$ & \cite{Lasinski-NPB-37-1-1972}& 16.26 & $11.30 \pm 0.10 \pm 0.17$ & \cite{Ayres-PRD-15-3105-1977}\\
2.768 & $7.20 \pm 0.20$ & & 5.757 & $9.79 \pm 0.40$ & \cite{Foley-PRL-11-425-1963}& 18.17 & $11.30 \pm 0.10 \pm 0.17$ & \\
2.768 & $7.80 \pm 0.15$ & \cite{Ambats-PRD-9-1179-1974}& 5.981 & $9.60 \pm 0.90$ & \cite{Lasinski-NPB-37-1-1972}& 19.46$^{*}$ & $12.64 \pm 0.12$ & \cite{Fajardo-PRD-24-46-1981}\\
2.972 & $8.29 \pm 0.16$ & & 6.028 & $9.79 \pm 0.63$ & \cite{Harting-NC-38-60-1965}& 27.60 & $10.70 \pm 0.90$ & \cite{Bromberg-PRL-31-1563-1973}\\
2.978 & $6.60 \pm 0.70$ & \cite{Lasinski-NPB-37-1-1972} & 6.028 & $9.96 \pm 0.21$ & & 31.00$^{*}$ & $12.83 \pm 0.34$ & \cite{Breakstone-NPB-248-253-1984}\\
3.077 & $8.20 \pm 1.00$ & & 6.059$^{*}$ & $12.50 \pm 0.70$ & \cite{Lasinski-NPB-37-1-1972}& 53.00 & $10.09 \pm 0.47$ & \\
3.363 & $8.46 \pm 0.16$ & \cite{Ambats-PRD-9-1179-1974}& 6.123 & $9.06 \pm 0.22$ & & 62.00 & $10.12 \pm 2.90$ & \\
3.497 & $8.70 \pm 0.50$ & \cite{Lasinski-NPB-37-1-1972} & 6.212 & $10.48 \pm 0.43$ & \cite{Foley-PRL-11-425-1963}&  &  & \\
3.627 & $8.46 \pm 0.16$ & \cite{Ambats-PRD-9-1179-1974}& 6.248 & $8.68 \pm 0.79$ & \cite{Foley-PRL-15-45-1965}&  &  & \\
\hline \multicolumn{9}{l}{$^*$\rule{0pt}{11pt}\footnotesize The
point is excluded from the fit procedure.}
\end{tabular}
\end{center}
\end{table*}

\begin{table*}[h!]
\caption{Experimental $B_{\bar{p}p}$ for low $|t|$ domain}
\label{tab:app3}
\begin{center}
\begin{tabular}{llc|llc|llc}
\hline \multicolumn{1}{c}{$\sqrt{s}$, GeV} &
\multicolumn{1}{c}{$B_{\bar{p}p}$, GeV$^{-2}$} &
\multicolumn{1}{c|}{Ref.} & \multicolumn{1}{c}{$\sqrt{s}$, GeV} &
\multicolumn{1}{c}{$B_{\bar{p}p}$, GeV$^{-2}$} &
\multicolumn{1}{c|}{Ref.} & \multicolumn{1}{c}{$\sqrt{s}$, GeV} &
\multicolumn{1}{c}{$B_{\bar{p}p}$, GeV$^{-2}$} &
\multicolumn{1}{c}{Ref.} \rule{0pt}{10pt}\\
\hline
1.885 & $83.70 \pm 24.00$ & \cite{Bruckner-PLB-158-180-1985} & 1.946 & $26.60 \pm 3.20$ & \cite{Lasinski-NPB-37-1-1972}& 2.638 & $13.10 \pm 0.40$ & \cite{Jenni-NPB-94-1-1975} \rule{0pt}{10pt}\\
1.887 & $129.0 \pm 25.00$ & \cite{Lasinski-NPB-37-1-1972} & 1.946 & $23.30 \pm 1.10$ & \cite{Cresti-PLB-132-209-1983}& 2.669 & $13.10 \pm 0.30$ &\cite{Lasinski-NPB-37-1-1972} \\
1.889 & $61.20 \pm 16.50$ & \cite{Bruckner-PLB-158-180-1985} & 1.948 & $22.00 \pm 2.00$ & \cite{Iwasaki-NPA-433-580-1985}& 2.957 & $14.10 \pm 1.60$ & \\
1.891 & $62.30 \pm 7.40$ & \cite{Lasinski-NPB-37-1-1972} & 1.949 & $19.12 \pm 0.83$ & \cite{Ashford-PRL-54-518-1985}& 2.987 & $12.90 \pm 0.39 \pm 0.08$ & \cite{Armstrong-PLB-385-479-1996}\\
1.891 & $52.40 \pm 3.80$ & \cite{Bruckner-PLB-158-180-1985} & 1.950 & $20.90 \pm 2.10$ & \cite{Schiavon-NPA-505-595-1989}& 3.077 & $13.50 \pm 0.90$ &\cite{Lasinski-NPB-37-1-1972} \\
1.891 & $71.50 \pm 4.50 \pm 7.00$ & \cite{Linssen-NPA-469-726-1987} & 1.951 & $25.90 \pm 1.70$ & \cite{Lasinski-NPB-37-1-1972}& 3.098 & $12.80 \pm 0.70 \pm 0.08$ & \cite{Armstrong-PLB-385-479-1996}\\
1.892 & $38.80 \pm 1.30$ & \cite{Iwasaki-NPA-433-580-1985} & 1.951 & $21.00 \pm 1.10$ & \cite{Cresti-PLB-132-209-1983}& 3.524 & $12.50 \pm 0.29 \pm 0.08$ & \\
1.894 & $39.20 \pm 2.00$ & \cite{Bruckner-PLB-158-180-1985} & 1.951 & $24.70 \pm 0.60$ & \cite{Iwasaki-NPA-433-580-1985}& 3.555 & $12.20 \pm 0.39 \pm 0.08$ & \\
1.896 & $51.20 \pm 4.60$ & \cite{Lasinski-NPB-37-1-1972} & 1.956 & $24.80 \pm 1.20$ & \cite{Lasinski-NPB-37-1-1972}& 3.612 & $12.60 \pm 0.29 \pm 0.08$ & \\
1.896 & $40.50 \pm 0.90$ & \cite{Iwasaki-NPA-433-580-1985} & 1.957 & $22.20 \pm 1.10$ & \cite{Cresti-PLB-132-209-1983}& 3.627 & $13.50 \pm 0.90$ & \cite{Jenni-NPB-129-232-1977}\\
1.896 & $47.70 \pm 2.70 \pm 5.00$ & \cite{Linssen-NPA-469-726-1987} & 1.957 & $24.10 \pm 0.60$ & \cite{Iwasaki-NPA-433-580-1985}& 3.686 & $12.20 \pm 0.59 \pm 0.08$ & \cite{Armstrong-PLB-385-479-1996}\\
1.898 & $32.30 \pm 1.70$ & \cite{Bruckner-PLB-158-180-1985} & 1.960 & $23.40 \pm 0.60$ & & 4.108 & $12.66 \pm 0.05$ & \cite{Lasinski-NPB-37-1-1972} \\
1.900 & $39.00 \pm 2.00$ & \cite{Iwasaki-NPA-433-580-1985} & 1.960 & $22.80 \pm 0.94$ & \cite{Bruckner-PLB-158-180-1985}& 4.108 & $13.00 \pm 0.50$ & \cite{Jenni-NPB-129-232-1977} \\
1.901 & $45.10 \pm 4.10$ & \cite{Lasinski-NPB-37-1-1972} & 1.962 & $24.90 \pm 1.20$ & \cite{Lasinski-NPB-37-1-1972}& 4.351 & $13.00 \pm 0.10$ & \cite{Gregory-NPB-119-60-1977}\\
1.902 & $35.70 \pm 0.80$ & \cite{Iwasaki-NPA-433-580-1985} & 1.962 & $18.79 \pm 0.80$ & \cite{Ashford-PRL-54-518-1985}& 4.540 & $11.80 \pm 3.00$ & \cite{Lasinski-NPB-37-1-1972}\\
1.906 & $42.20 \pm 4.80$ & \cite{Lasinski-NPB-37-1-1972} & 1.969 & $18.23 \pm 0.80$ & & 4.540 & $12.20 \pm 0.10$ & \cite{Jenni-NPB-129-232-1977}\\
1.907 & $36.50 \pm 0.80$ & \cite{Iwasaki-NPA-433-580-1985} & 1.973 & $20.40 \pm 2.20$ & \cite{Lasinski-NPB-37-1-1972}& 4.621 & $11.38 \pm 0.17 \pm 0.10$ & \cite{Carnegie-PLB-59-313-1975} \\
1.908 & $32.10 \pm 3.50$ & \cite{Lasinski-NPB-37-1-1972} & 1.976 & $21.73 \pm 0.85$ & \cite{Ashford-PRL-54-518-1985}& 11.54 & $12.49 \pm 0.18 \pm 0.33$ & \cite{Fajardo-PRD-24-46-1981}\\
1.908 & $33.50 \pm 1.20$ & \cite{Cresti-PLB-132-209-1983} & 1.985 & $15.90 \pm 0.80$ & \cite{Iwasaki-NPA-433-580-1985}& 13.76 & $13.20 \pm 1.20$ & \cite{Cool-PRD-24-2821-1981}\\
1.909 & $44.15 \pm 5.55$ & \cite{Ashford-PRL-54-518-1985} & 1.989 & $15.00 \pm 1.80$ & \cite{Kaseno-PLB-61-203-1976}& 15.36 & $12.46 \pm 0.14 \pm 0.33$ & \cite{Fajardo-PRD-24-46-1981}\\
1.911 & $28.60 \pm 6.80$ & \cite{Lasinski-NPB-37-1-1972} & 1.994 & $12.00 \pm 2.00$ & \cite{Iwasaki-NPA-433-580-1985}& 16.91 & $12.52 \pm 0.16 \pm 0.33$& \\
1.912 & $33.60 \pm 1.70$ & \cite{Iwasaki-NPA-433-580-1985} & 1.996 & $14.10 \pm 2.60$ & \cite{Lasinski-NPB-37-1-1972}& 18.14 & $13.03 \pm 0.19 \pm 0.33$ & \\
1.917 & $27.40 \pm 1.70$ & \cite{Ashford-PRL-54-518-1985} & 2.005 & $16.30 \pm 1.20$ & & 19.42 & $17.00 \pm 3.60$ & \cite{Cool-PRD-24-2821-1981}\\
1.918 & $24.80 \pm 2.60$ & \cite{Lasinski-NPB-37-1-1972} & 2.006 & $18.00 \pm 0.50$ & \cite{Schiavon-NPA-505-595-1989}& 19.46 & $12.68 \pm 0.10 \pm 0.32$ & \cite{Fajardo-PRD-24-46-1981}\\
1.918 & $28.80 \pm 1.20$ & \cite{Cresti-PLB-132-209-1983} & 2.036 & $18.60 \pm 1.50$ & \cite{Lasinski-NPB-37-1-1972}& 24.30 & $12.30 \pm 0.50 \pm 0.07$ & \cite{Breedon-PLB-216-459-1989} \\
1.920 & $29.10 \pm 0.50$ & \cite{Iwasaki-NPA-433-580-1985} & 2.107 & $15.20 \pm 0.30$ & \cite{Schiavon-NPA-505-595-1989}& 30.40 & $12.70 \pm 0.50$ & \cite{Amos-NPB-262-689-1985} \\
1.921 & $22.70 \pm 4.20$ & \cite{Lasinski-NPB-37-1-1972} & 2.115 & $16.10 \pm 0.40$ & \cite{Iwasaki-NPA-433-580-1985} & 30.70 & $12.60 \pm 0.30$ & \cite{Amos-PLB-128-343-1983} \\
1.922 & $24.70 \pm 1.70$ & \cite{Iwasaki-NPA-433-580-1985} & 2.138 & $15.00 \pm 0.30$ & & 31.00 & $11.37 \pm 0.60$ & \cite{Breakstone-NPB-248-253-1984} \\
1.923 & $22.63 \pm 1.30$ & \cite{Ashford-PRL-54-518-1985} & 2.140 & $14.90 \pm 0.60$ & \cite{Jenni-NPB-94-1-1975}& 52.60 & $13.03 \pm 0.52$ & \cite{Amos-NPB-262-689-1985}\\
1.926 & $23.80 \pm 2.30$ & \cite{Lasinski-NPB-37-1-1972} & 2.159 & $14.20 \pm 0.30$ & \cite{Iwasaki-NPA-433-580-1985}& 52.80 & $13.60 \pm 2.20$ & \cite{Favart-PRL-47-1191-1981}\\
1.926 & $26.80 \pm 1.20$ & \cite{Cresti-PLB-132-209-1983} & 2.184 & $15.20 \pm 0.30$ & & 52.80 & $13.92 \pm 0.37 \pm 0.22$ & \cite{Ambrosio-PLB-115-495-1982}\\
1.926 & $26.20 \pm 0.50$ & \cite{Iwasaki-NPA-433-580-1985} & 2.193 & $15.40 \pm 0.30$ & & 52.80 & $13.36 \pm 0.53$ & \cite{Amos-PLB-120-460-1983} \\
1.927 & $23.38 \pm 1.15$ & \cite{Ashford-PRL-54-518-1985} & 2.204 & $14.50 \pm 0.30$ & & 62.30 & $13.47 \pm 0.52$ & \cite{Amos-NPB-262-689-1985}\\
1.931 & $22.30 \pm 1.50$ & \cite{Lasinski-NPB-37-1-1972} & 2.221 & $14.20 \pm 0.20$ & & 62.50 & $13.10 \pm 0.60$ & \cite{Amos-PLB-128-343-1983}\\
1.931 & $23.70 \pm 0.40$ & \cite{Iwasaki-NPA-433-580-1985} & 2.223 & $14.20 \pm 0.30$ & \cite{Jenni-NPB-94-1-1975}& 540.0 & $17.20 \pm 1.00$ & \cite{Battiston-PLB-115-333-1982}\\
1.932 & $22.52 \pm 1.05$ & \cite{Ashford-PRL-54-518-1985} & 2.227 & $15.80 \pm 0.30$ & \cite{Iwasaki-NPA-433-580-1985}& 540.0 & $17.60 \pm 1.00$ & \cite{Battiston-PLB-127-472-1983} \\
1.933 & $23.80 \pm 1.20$ & \cite{Cresti-PLB-132-209-1983} & 2.254$^{*}$ & $5.20 \pm 0.30$ & & 540.0 & $17.10 \pm 1.00$ & \cite{Arnison-PLB-128-336-1983}\\
1.936 & $21.80 \pm 0.60$ & \cite{Iwasaki-NPA-433-580-1985} & 2.260& $16.50 \pm 0.40$ & & 541.0 & $15.50 \pm 0.20$ & \cite{Augier-PLB-316-448-1993} \\
1.938 & $18.80 \pm 1.90$ & \cite{Lasinski-NPB-37-1-1972} & 2.292 & $15.10 \pm 1.80$ & \cite{Lasinski-NPB-37-1-1972} & 546.0 & $15.30 \pm 0.30$ & \cite{Bozzo-PLB-147-385-1984}\\
1.938 & $18.98 \pm 0.90$ & \cite{Ashford-PRL-54-518-1985} & 2.294 & $12.40 \pm 0.50$ & \cite{Iwasaki-NPA-433-580-1985}& 546.0 & $15.20 \pm 0.20$ & \\
1.939 & $26.40 \pm 1.14$ & \cite{Bruckner-PLB-158-180-1985} & 2.296 & $14.40 \pm 0.40$ & \cite{Lasinski-NPB-37-1-1972}& 546.0 & $15.28 \pm 0.28 \pm 0.09$ & \cite{Abe-PRD-50-5518-1994}\\
1.940 & $25.20 \pm 1.10$ & \cite{Cresti-PLB-132-209-1983} & 2.304$^{*}$& $5.90 \pm 0.30$ & \cite{Iwasaki-NPA-433-580-1985} & 1800 & $17.20 \pm 1.30$ & \cite{Amos-PRL-61-525-1988}\\
1.942 & $22.40 \pm 2.70$ & \cite{Lasinski-NPB-37-1-1972} & 2.325 & $13.30 \pm 1.30$ & \cite{Lasinski-NPB-37-1-1972} & 1800 & $16.30 \pm 0.50$ & \cite{Amos-PRL-63-2784-1989} \\
1.943 & $20.01 \pm 0.82$ & \cite{Ashford-PRL-54-518-1985} & 2.348 & $13.30 \pm 0.20$ & \cite{Iwasaki-NPA-433-580-1985}& 1800 & $16.37 \pm 0.45$ & \cite{Amos-PLB-247-127-1990}\\
1.944 & $20.10 \pm 2.20$ & \cite{Lasinski-NPB-37-1-1972} & 2.352 & $13.20 \pm 0.30$ & \cite{Jenni-NPB-94-1-1975}& 1800 & $16.99 \pm 0.47$ & \cite{Amos-PRL-68-2433-1992} \\
1.945 & $22.80 \pm 0.40$ & \cite{Iwasaki-NPA-433-580-1985} & 2.430 & $15.60 \pm 1.40$ & \cite{Lasinski-NPB-37-1-1972}& & & \\
\hline \multicolumn{9}{l}{$^*$\rule{0pt}{11pt}\footnotesize The
point is excluded from the fit procedure.}
\end{tabular}
\end{center}
\end{table*}

\begin{table*}[h!]
\caption{Experimental $B_{\bar{p}p}$ for intermediate $|t|$
domain} \label{tab:app4}
\begin{center}
\begin{tabular}{llc|llc|llc}
\hline \multicolumn{1}{c}{$\sqrt{s}$, GeV} &
\multicolumn{1}{c}{$B_{\bar{p}p}$, GeV$^{-2}$} &
\multicolumn{1}{c|}{Ref.} & \multicolumn{1}{c}{$\sqrt{s}$, GeV} &
\multicolumn{1}{c}{$B_{\bar{p}p}$, GeV$^{-2}$} &
\multicolumn{1}{c|}{Ref.} & \multicolumn{1}{c}{$\sqrt{s}$, GeV} &
\multicolumn{1}{c}{$B_{\bar{p}p}$, GeV$^{-2}$} &
\multicolumn{1}{c}{Ref.} \rule{0pt}{10pt}\\
\hline
\multicolumn{9}{c}{Linear parameterization $\ln(d\sigma / dt) \propto (-B|t|)$} \rule{0pt}{10pt}\\
\hline
2.156$^{*}$ & $17.80 \pm 0.40$ & \cite{Lasinski-NPB-37-1-1972} & 4.307 & $12.83 \pm 0.21$ & \cite{Lasinski-NPB-37-1-1972} & 31.00 & $11.16 \pm 0.20$ & ~\cite{Breakstone-NPB-248-253-1984} \rule{0pt}{10pt}\\
2.177 & $11.80 \pm 1.00$ & & 4.540 & $11.80 \pm 2.90$ & \cite{Foley-PRL-11-503-1963} & 52.80 & $10.68 \pm 0.20$& \cite{Ambrosio-PLB-115-495-1982} \\
2.271 & $11.20 \pm 1.30$ & & 4.621 & $12.09 \pm 0.24 \pm 0.10$ & \cite{Carnegie-PLB-59-313-1975} & 53.00 & $11.50 \pm 0.15$ & \cite{Breakstone-NPB-248-253-1984}\\
2.335 & $16.50 \pm 1.00$ & & 4.896 & $12.33 \pm 0.79$ & \cite{Foley-PRL-15-45-1965} & 62.00 & $11.12 \pm 0.15$ &\\
2.465 & $12.90 \pm 0.70$ & & 4.896 & $12.30 \pm 0.80$ & \cite{Lasinski-NPB-37-1-1972} & 540.0 & $13.30 \pm 1.50$& \cite{Arnison-PLB-121-77-1983}\\
2.558 & $13.40 \pm 1.10$ & & 4.934 & $12.66 \pm 0.29$ & \cite{Foley-PRL-11-503-1963} & 540.0 & $13.70 \pm 0.30$& \cite{Battiston-PLB-127-472-1983}\\
2.702 & $13.00 \pm 2.30$ & & 4.934 & $12.70 \pm 0.30$ & \cite{Lasinski-NPB-37-1-1972} & 540.0 & $13.70 \pm 0.20 \pm 0.20$& \cite{Arnison-PLB-128-336-1983} \\
2.768 & $13.00 \pm 0.80$ & & 5.627$^{*}$ & $8.78 \pm 1.00$ & \cite{Foley-PRL-15-45-1965} & 546.0 & $13.40 \pm 0.30$ & \cite{Bozzo-PLB-147-385-1984} \\
2.857 & $13.10 \pm 0.70$ & & 5.642$^{*}$ & $11.44 \pm 0.20$ & \cite{Birnbaum-PRL-23-663-1969} & 546.0 & $13.60 \pm 0.80$ &\\
3.550 & $12.10 \pm 0.40$ & & 5.642$^{*}$ & $11.64 \pm 0.08$ & \cite{Lasinski-NPB-37-1-1972} & 546.0 & $14.20 \pm 0.40$ &\\
3.550 & $12.60 \pm 0.20$ & \cite{Braun-NPB-95-481-1975} & 6.621 & $13.10 \pm 0.80$ & \cite{Batyunya-YaF-44-1489-1986} & 546.0$^{*}$& $15.35 \pm 0.18 \pm 0.06$ & \cite{Abe-PRD-50-5518-1994} \\
3.851$^{*}$ & $15.60 \pm 0.40$ & \cite{Lasinski-NPB-37-1-1972} & 7.006 & $11.80 \pm 0.10 \pm 0.12$ & \cite{Antipov-NPB-57-333-1973} & 1020 & $16.20 \pm 0.50 \pm 0.50$ & \cite{Amos-NCA-106-123-1993} \\
3.923 & $13.15 \pm 0.47$ & \cite{Foley-PRL-11-503-1963} & 7.864 & $13.26 \pm 0.30$ & \cite{Bogolyubsky-YaF-41-1210-1984} & 1800 & $16.30 \pm 0.30$ & \cite{Amos-PLB-247-127-1990} \\
3.923 & $13.10 \pm 0.40$ & \cite{Lasinski-NPB-37-1-1972} & 7.875 & $12.70 \pm 0.80$ & \cite{Grard-PLB-59-409-1975} & 1800 & $16.40 \pm 0.12$ &\\
4.108 & $12.57 \pm 0.20$ & \cite{Birnbaum-PRL-23-663-1969} & 8.777 & $11.30 \pm 0.20 \pm 0.11$ & \cite{Antipov-NPB-57-333-1973} & 1800 & $16.98 \pm 0.24 \pm 0.07$ & \cite{Abe-PRD-50-5518-1994} \\
4.108 & $12.72 \pm 0.11$ & \cite{Lasinski-NPB-37-1-1972} & 11.54 & $10.10 \pm 1.00$ & \cite{Dumont-ZPC-13-1-1982}& 1960 & $16.86 \pm 0.10 \pm 0.20$ & \cite{Abazov-PRD-86-012009-2012} \\
4.307 & $12.84 \pm 0.21$ & \cite{Foley-PRL-11-503-1963} & 13.76 & $11.40 \pm 0.60$ & \cite{Ansorge-PLB-59-299-1975} & & &\\
\hline
\multicolumn{9}{c}{Quadratic parameterization $\ln(d\sigma / dt) \propto (-B|t| \pm Ct^{2})$} \rule{0pt}{10pt}\\
\hline
2.768 & $12.20 \pm 0.80$ & \cite{Ambast-PRD-9-1179-1974} & 8.777 & $12.20 \pm 0.70 \pm 0.12$ & \cite{Antipov-NPB-57-333-1973}& 31.00 & $13.06 \pm 0.39$ & ~\cite{Breakstone-NPB-248-253-1984} \rule{0pt}{10pt}\\
2.972 & $12.10 \pm 1.00$ & & 9.778 & $12.60 \pm 0.20 \pm 0.19$ & \cite{Ayres-PRD-15-3105-1977}& 53.00$^{*}$& $10.96 \pm 0.44$ & \\
3.363 & $11.40 \pm 1.00$ & & 11.54 & $12.80 \pm 0.30 \pm 0.19$ & & 62.00$^{*}$ & $11.30 \pm 0.45$ & \\
3.627 & $12.40 \pm 1.50$ & & 13.76 & $11.90 \pm 0.50 \pm $ 0.18& & 1800 & $16.25 \pm 0.23$ & \cite{Amos-PLB-247-127-1990} \\
6.621 & $13.10 \pm 0.80$ & \cite{Batyunya-YaF-44-1489-1986}& 16.26 & $12.60 \pm 0.40 \pm 0.19$ & & 1800 & $16.87 \pm 0.57$ & \\
7.006 & $12.80 \pm 0.40 \pm 0.13$ & \cite{Antipov-NPB-57-333-1973}& 18.17 & $13.10 \pm 0.40 \pm 0.20$ & & 1800 & $20.65 \pm 3.84$ & \\
7.864 & $13.26 \pm 0.30$ & \cite{Bogolyubsky-YaF-41-1210-1984}& 19.46$^{*}$ & $13.27 \pm 0.24$ & \cite{Fajardo-PRD-24-46-1981} & & & \\
\hline \multicolumn{9}{l}{$^*$\rule{0pt}{11pt}\footnotesize The
point is excluded from the fit procedure.}
\end{tabular}
\end{center}
\end{table*}

\end{document}